\documentclass[%
preprint,
amsmath,amssymb,
aps,
]{revtex4-2}
\usepackage[utf8]{inputenc}
\usepackage[T1]{fontenc} 
\usepackage{amsfonts}
\usepackage{amsthm}
\usepackage{booktabs}
\usepackage{cancel}
\usepackage{bbm}
\usepackage{slashed}
\usepackage{color}
\usepackage[dvipsnames]{xcolor}
\usepackage{empheq}
\usepackage{graphicx}
\usepackage{hyperref}
\usepackage[normalem]{ulem}
\usepackage[margin=0.9in]{geometry}
\newcommand\fig[1] {Fig.\,{\ref{#1}}}
\newcommand\sect[1] {Sect.\,{\ref{#1}}}

\newcommand\app[1] {Appendix~\ref{#1}}

\newcommand{\be}{\begin{equation}}
    \newcommand{\ee}{\end{equation}}
\newcommand{\beq}{\begin{equation}}
    \newcommand{\eeq}{\end{equation}}
\def\bsp#1\esp{\begin{split}#1\end{split}}
\def\bal#1\eal{\begin{align}#1\end{align}}

\newcommand\bom[1]     {{\mbox{\boldmath $#1$}}}

\newcommand{\rL}{\mathrm{L}}
\newcommand{\tS}{\theta_S}
\newcommand{\cL}{\mathcal{L}}
\newcommand{\half}{\frac{1}{2}}
\newcommand{\hc}{\text{h.c.}}

\newcommand{\SARAH}{{\tt SARAH}}

\def\bal#1\eal{\begin{align}#1\end{align}}
\newcommand{\ri}{\mathrm{i}}
\newcommand{\rd}{\mathrm{d}}
\newcommand{\rf}{\mathrm{f}}
\newcommand{\rs}{\mathrm{s}}
\newcommand{\rE}{\mathrm{E}}
\newcommand{\rO}{\mathrm{O}}
\newcommand{\gL}{g_{\rm L}}
\newcommand{\tG}{\theta_{\rm G}}
\newcommand{\tT}{\theta_Z}
\newcommand{\tW}{\theta_{\rm W}}
\newcommand{\nub}{\nu}
\newcommand{\sgn}[1]{\operatorname{sgn}\left(#1\right)}
\newcommand{\brac}[1]{\left( #1 \right)}
\newcommand{\cl}{\cancel{\ell}}
\newcommand{\bg}{\bom{\Gamma}}

\numberwithin{equation}{section}

\begin{document}
    
    \title{One-loop corrections to light neutrino masses in gauged U(1) extensions of the standard model}
    
    \author{Sho Iwamoto}
    \email{sho.iwamoto@ttk.elte.hu}
    \affiliation{Institute for Theoretical Physics, ELTE E\"otv\"os Lor\'and University,
        P\'azm\'any P\'eter s\'et\'any 1/A, 1117 Budapest, Hungary}
    \author{Timo J.~K\"arkk\"ainen}
    \email{timo.karkkainen@ttk.elte.hu}
    \affiliation{Institute for Theoretical Physics, ELTE E\"otv\"os Lor\'and University,
        P\'azm\'any P\'eter s\'et\'any 1/A, 1117 Budapest, Hungary}
    \author{Zolt\'an P\'eli}
    \email{zoltanpeli92@gmail.com}
    \affiliation{ELKH-DE Particle Physics Research Group, 4010 Debrecen, PO Box 105, Hungary}
    \author{Zolt\'an Tr\'ocs\'anyi}
    \email{zoltan.trocsanyi@cern.ch}
    \affiliation{Institute for Theoretical Physics, ELTE E\"otv\"os Lor\'and University,
        P\'azm\'any P\'eter s\'et\'any 1/A, 1117 Budapest, Hungary\\
        ELKH-DE Particle Physics Research Group, 4010 Debrecen, PO Box 105, Hungary}
    \begin{abstract}
        We consider gauged U(1) extensions of the standard model of particle physics with
        three right-handed sterile neutrinos and a singlet scalar. The neutrinos obtain 
        mass via the type I seesaw mechanism. We compute the one loop corrections to the
        elements of the tree level mass matrix of the light neutrinos and show explicitly 
        the cancellation of the gauge dependent terms. We present a general formula for 
        the gauge independent, finite one-loop corrections for arbitrary number new U(1) 
        groups, new complex scalars and sterile neutrinos. We estimate the size of the
        corrections relative to the tree level mass matrix in a particular extension, the 
        super-weak model.
    \end{abstract}
    \keywords{gauged U(1) extension, neutrino mass, seesaw mechanism, gauge independence}
    \maketitle
    \thispagestyle{empty}
    
    \section{Introduction}
    The standard model (SM) of particle interactions is one of the most successful
    physics models with unprecedented precision for predicting physical quantities, for 
    instance for the anomalous magnetic moment of the electron. However, it does not
    contain right handed neutrinos as they are sterile under the SM gauge group. 
    This in turn leads to the prediction that neutrinos are massless which is in
    conflict of the now well established experimental result, that at least two 
    neutrinos are massive \cite{SK98,SNO01}, and therefore, signals that the SM 
    requires an extension to explain the origin of the neutrino masses. There are 
    lots of models attempting to explain neutrino masses. Among those perhaps 
    the most economical one that requires the least extension of the SM, is the 
    type I seesaw mechanism where neutrinos acquire masses after spontaneous 
    symmetry breaking (SSB) of one or more scalar fields
    \cite{Fritzsch75,Minkowski:1977sc,GellMann:1980vs,Yanagida:1979as,Mohapatra:1979ia,Schechter:1980gr,Magg:1980ut,Glashow:1979nm}.
    
    Recently there has been a lot of interest in gauged U(1)-extended models in 
    particle physics phenomenology, motivated by the observed difference 
    between the measured and SM predicted values of the anomalous magnetic
    moment of the muon \cite{Abi:2021gix} and also anomalies in short-baseline 
    neutrino oscillations \cite{Aguilar-Arevalo:2018gpe}. Gauged $B-L$, $B-3L_\tau$,
    $L_e-L_\mu$ and $L_\mu-L_\tau$ have been considered \cite{Nelson:2007yq,Heeck:2010pg,
        Ma:1997nq,Ma:2001md,Asai:2019ciz,Bhatia:2017tgo,Adhikari:2008uc}, 
    as well as a general gauged U(1) not related to flavour \cite{Borah:2020swo}. 
    In these models, both seesaw and radiative one-loop neutrino mass generation 
    mechanisms have been considered.
    
    As the effects of new physics are typically much smaller than those of the SM
    interactions, computations in theories beyond the SM are often considered only at
    tree level. Yet, the loop corrections may be sizable and can affect significantly 
    the validity region in the parameter space of the model. For instance, the 
    lightness of active neutrinos requires that the loop corrections to the mass 
    matrix of those particles must also be small in order to have a phenomenologically 
    viable model. Computations of such one-loop corrections have been carried out 
    previously in Refs.~\cite{AristizabalSierra:2011mn,LopezPavon:2012zg} for the
    canonical seesaw case, and in the context of multi-Higgs doublet models
    \cite{Grimus:2002nk,Grimus:2018rte,Dudenas:2018wlr,Denner:2016etu}.
    In the cases of gauged U(1) models we are not aware of a computation
    of the \textit{one-loop corrections} to active neutrino mass matrix.
    
    In this article we consider gauged U(1) extensions of the SM and derive a general 
    formula for the one-loop corrections of the mass matrix of the active neutrinos.
    The mass matrix of the active neutrinos emerges after SSB due to the type I seesaw 
    mechanism. Our goal is to derive the one-loop corrections to that mass matrix and 
    estimate their sizes relative to the tree level for a particular example, called 
    the super-weak force \cite{Trocsanyi:2018bkm}. The super-weak model contains three
    additional right-handed sterile (under the SM interactions) neutrinos and one 
    complex scalar field in addition to the fields of the SM. The loop corrections 
    involve all the gauge and scalar bosons which couple to neutrinos.
    
    In order to obtain the one-loop corrections to the elements of the light
    neutrino mass matrix, we perform our computations in the $R_\xi$ 
    gauge and show explicitly the intricate cancellation of the gauge fixing 
    parameters from the corrections. In addition, we shall also demonstrate the 
    cancellation of the $\epsilon$ poles when the loop 
    integrals are regulated by dimensional regularization in $d=4-2\epsilon$ 
    dimensions. These cancellations are highly non-trivial, and therefore provide
    strong checks on the correctness of the computations.
    
    The paper is composed as follows. We introduce the model to the extent 
    needed for the present work in \sect{sec:Notation}. We define and 
    compute the one loop correction to mass matrix of the active neutrinos 
    in \sect{sec:1loop}. In \sect{sec:numerics} we provide numerical estimates 
    of the one-loop corrections and show that those are very small.
    Finally we summarize our findings in Sec.~\ref{sec:conclusions}. 
    We collect auxiliary formulas in the appendices and also provide an auxiliary 
    zip file containing the \SARAH\ {\tt model}, {\tt parameter} and {\tt particle} files.
    
    \section{Particle model, mixings and interactions}
    \label{sec:Notation}
    
    We consider an extension of the standard model by a U(1)$_z$ gauge group with
    particle content and charge assignment defined in Ref.~\cite{Trocsanyi:2018bkm}.
    The {\em super-weak model} is an economical extension of the standard model 
    that provides a framework to explain the origin of
    (i) neutrino masses and oscillations \cite{Karkkainen:2021}, 
    (ii) dark matter \cite{Iwamoto:2021fup}, 
    (iii) cosmic inflation and stabilization of the electroweak vacuum \cite{Peli:2019vtp},
    (iv) matter-antimatter asymmetry of the universe. The complete model 
    including Feynman rules in the unitary gauge was presented fully in
    Ref.~\cite{Trocsanyi:2018bkm}. As we are to 
    compute one-loop corrections to neutrino masses, we recall the details
    relevant to such computations, with Feynman rules in the $R_\xi$
    gauge. We generated those Feynman rules with \SARAH
    \cite{Staub:2008uz,Staub:2009bi,Staub:2010jh,Staub:2013tta}
    but here we present simpler forms for the rules needed in our
    computations to make those more comprehensive. We also recall some of
    the conventions that are different in \SARAH\ and the original
    definition of the model. We stick to the \SARAH\ conventions 
    throughout this work.%
    \footnote{We present the model files in a separate file {\tt SuperWeak.zip}.}
    
    \subsection{Mixing of neutral gauge bosons}
    \label{sec:AZTmixing}
    
    The particle content of the standard model is extended by 3 right-handed 
    neutrinos $\nu_\mathrm{R}$, a new scalar $\chi$, and the U(1)$_z$ gauge 
    boson $B'$. As the field strength tensors of the U(1) gauge groups are 
    gauge invariant, kinetic mixing is allowed between the gauge fields 
    belonging to the hypercharge U(1)$_y$ and the new U(1)$_z$ gauge symmetries, 
    whose strength is measured by $\epsilon$ in
    \be
    \bsp
    \mathcal L &\supset
    -\frac14 F^{\mu\nu} F_{\mu\nu}
    -\frac14 F'^{\mu\nu} F'_{\mu\nu}
    -\frac{\epsilon}{2} F^{\mu\nu} F'_{\mu\nu}
    \,,\\
    \mathcal{D}^{\mathrm{U}(1)}_\mu &= -\ri
    (y g_y B_\mu + z g_z B'_\mu)
    \label{eq:kineticmixing}
    \esp
    \ee
    where $B^\mu$ is the U(1)$_y$ gauge field. However, equivalently, we can 
    choose the basis---the convention in  \texttt{SARAH}---%
    in which the gauge-field strengths do not mix, while the couplings are given 
    by a $2\times 2$ coupling matrix in the covariant derivative
    \begin{equation}
        D^{\mathrm{U}(1)}_\mu = -\ri \begin{pmatrix}
            y & z
        \end{pmatrix}
        \begin{pmatrix}
            \hat g_{yy} & \hat g_{yz} \\
            \hat g_{zy} & \hat g_{zz}
        \end{pmatrix}
        \begin{pmatrix}
            \hat B_\mu \\ \hat B'_\mu
        \end{pmatrix}
    \end{equation}
    where $y$ and $z$ are the U(1)$_y$ and U(1)$_z$ charges. We can parametrize
    the coupling matrix as
    \begin{equation}
        \hat{\textbf{g}}=\begin{pmatrix}
            \hat g_{yy} & \hat g_{yz} \\
            \hat g_{zy} & \hat g_{zz}
        \end{pmatrix} = 
        \begin{pmatrix}
            g_y & -\eta g'_z \\
            0 & g'_z
        \end{pmatrix}
        \begin{pmatrix}
            \cos\epsilon' & \sin\epsilon' \\
            -\sin\epsilon' & \cos\epsilon'
        \end{pmatrix}\,.
    \end{equation}
    The coupling mixing matrix containing $\eta$ is equivalent to the kinetic mixing 
    in the Lagrangian \eqref{eq:kineticmixing} and the parameters of the two 
    representations are related by $g'_z = g_z/\sqrt{1-\epsilon^2}$ and 
    $\eta = \epsilon g_y/g_z$. In this paper, it will be convenient to use the 
    kinetic mixing representation defined by \eqref{eq:kineticmixing}.
    
    The rotation with angle $\epsilon'$ is unphysical as it can be
    absorbed into the mixing of the neutral gauge fields $B^\mu$, $B'^\mu$
    and $W^{3\,\mu}$ to the mass eigenstates $A^\mu$, $Z^\mu$ and
    $Z^{\prime\mu}$, which then can be described by a rotation matrix
    \beq
    \left(\begin{array}{c}
        \hat B^\mu \\
        W^{3\,\mu} \\
        \hat B'^\mu
    \end{array}\right) =
    \left( \begin{array}{ccc}
        \cos\tW  & - \cos\tT  \sin\tW   & - \sin\tT
        \sin\tW   \\
        \sin\tW  & \cos\tT  \cos\tW   & \cos\tW  \sin\tT
        \\
        0 & - \sin\tT   & \cos\tT \end{array}
    \right)\left(\begin{array}{c}
        A^\mu \\
        Z^\mu \\
        Z^{\prime\mu}
    \end{array}\right).
    \eeq
    This matrix depends on two mixing angles: $\tW$ is the weak mixing 
    (or Weinberg) angle and $\tT$ is the $Z-Z'$ mixing angle
    \footnote{Note the opposite sign convention for $\tT$ in this work and 
        in Ref.~\cite{Trocsanyi:2018bkm} where this mixing angle was 
        denoted as $\theta_T$, so $\tT = -\theta_T$.}.
    In terms of the coupling parameters 
    \beq
    \kappa = \cos\tW(\gamma'_y-2\gamma_z')
    \text{~~and~~}
    \tau = 2 \cos\tW \gamma_z' \tan \beta
    \,,
    \label{eq:kappa-tau}
    \eeq
    introduced in Ref.~\cite{Trocsanyi:2018bkm}, this new mixing angle is given
    implicitly by $\tan(2 \tT) = 2 \kappa/(1 - \kappa^2 - \tau^2)$. In 
    Eq.~\eqref{eq:kappa-tau} $\tan\beta = w/v$ is the ratio of the vacuum expectation 
    values (VEVs) of the scalar fields (see below) and 
    $\gamma'_y = (\epsilon/\sqrt{1-\epsilon^2})(g_y/\gL)$, $\gamma'_z = g'_z/\gL$, 
    i.e.~the couplings are normalized by the
    SU(2)$_{\rm L}$ coupling.
    
    We can express the elements of the $Z-Z'$ mixing matrix explicitly,
    \beq
    \bsp
    \sin \tT &= \sgn{\kappa}\left[\frac12\left(1
    - \frac{1 - \kappa^2 - \tau^2}{\sqrt{(1 + \kappa^2 + \tau^2)^2-4\tau^2}}
    \right)\right]^{1/2}
    ,\\
    \cos \tT &= \left[\frac12\left(1
    + \frac{1 - \kappa^2 - \tau^2}{\sqrt{(1 + \kappa^2 + \tau^2)^2-4\tau^2}}
    \right)\right]^{1/2}
    \,,
    \label{eq:sT-cT}
    \esp
    \eeq
    which also appear in the neutral currents $\Gamma^\mu_{V\bar{f}f} =
    -\ri e \gamma^\mu (C_{V\bar{f}f}^R P_R + C_{V\bar{f}f}^L P_L)$
    where $e$ is the electromagnetic coupling and
    $P_{R/L} \equiv P_\pm = \half (1\pm \gamma^5)$ are the usual chiral
    projections. In particular, for neutrinos
    \beq
    \bsp
    eC_{Z\nu\nu}^L &= \frac{\gL}{2\cos\tW}
    \Big[\cos \tT - (\gamma_y'-\gamma_z') \sin \tT \cos\tW \Big]
    \,,\quad
    eC_{Z\nu\nu}^R = -\frac{\gL}{2} \gamma_z' \sin \tT
    \,,\\
    eC_{Z'\nu\nu}^L &=  \frac{\gL}{2\cos\tW}
    \Big[\sin \tT + (\gamma_y'-\gamma_z') \cos \tT \cos\tW \Big]
    \,,\quad
    eC_{Z'\nu\nu}^R = \frac{\gL}{2} \gamma_z' \cos \tT
    \,,
    \label{eq:eCLR}
    \esp
    \eeq
    i.e. $C_{Z'\nu\nu}^{L/R}$ can be obtained from $C_{Z\nu\nu}^{L/R}$ by
    the replacement
    \beq
    (Z\to Z') \Rightarrow (\cos \tT, \sin \tT)\to (\sin \tT, -\cos \tT)
    \,.
    \label{eq:ZtoT}
    \eeq
    
    \subsection{Mixings of scalar and Goldstone bosons}
    \label{sec:Gmixing}
    
    In addition to the usual $SU(2)_\rL$-doublet Brout-Englert-Higgs (BEH) field
    \begin{equation}
        \phi=\left(\!\!\begin{array}{c}
            \phi^{+} \\
            \phi^{0}
        \end{array}\!\!\right) = 
        \frac{1}{\sqrt{2}}
        \left(\!\!\begin{array}{c}
            \phi_{1}+\ri\phi_{2} \\
            \phi_{3}+\ri\phi_{4}
        \end{array}\!\!
        \right)
        \,,
    \end{equation}
    there is another complex scalar $\chi$ in the model, with charges specified in
    \cite{Trocsanyi:2018bkm}.
    The Lagrangian of the scalar fields contains the potential energy
    \beq
    V(\phi,\chi) = V_0 - \mu_\phi^2 |\phi|^2 - \mu_\chi^2 |\chi|^2
    + \left(|\phi|^2, |\chi|^2\right)
    \left(\!\!\begin{array}{cc}
        \lambda_\phi & \frac{\lambda}2 \\ \frac{\lambda}2 & \lambda_\chi
    \end{array}\!\!\right)
    \left(\!\!\begin{array}{c}
        |\phi|^2 \\ |\chi|^2
    \end{array}\!\!\right) \subset  -\cL
    \label{eq:V}
    \eeq
    where $|\phi|^2 =|\phi^+|^2 + |\phi^0|^2$.
    In the $R_\xi$ gauge we parametrize the scalar fields after
    spontaneous symmetry breaking as
    \beq
    \bsp
    \phi =\frac{1}{\sqrt{2}}\binom{-\ri \sqrt{2}\sigma^+}{v+h'+ \ri\sigma_\phi}
    \,,\quad
    \chi = \frac{1}{\sqrt{2}}(w + s' + \ri\sigma_\chi)
    \esp
    \eeq
    where
    $v$ and $w$ denotes the vacuum expectation values (VEVs) of the
    fields, whose values are
    \beq
    v = \sqrt{2} \sqrt{\frac{2\lambda_\chi \mu_\phi^2 - \lambda \mu_\chi^2}
        {4 \lambda_\phi \lambda_\chi - \lambda^2}}
    \,,\qquad
    w = \sqrt{2} \sqrt{\frac{2\lambda_\phi \mu_\chi^2 - \lambda \mu_\phi^2}
        {4 \lambda_\phi \lambda_\chi - \lambda^2}}
    \,.
    \label{eq:VEVs}
    \eeq
    Using the VEVs, we can express the quadratic couplings as
    \beq
    \mu_\phi^2 = \lambda_\phi v^2 + \frac{\lambda}{2} w^2
    \,,\qquad
    \mu_\chi^2 = \lambda_\chi w^2 + \frac{\lambda}{2} v^2
    \,.
    \label{eq:scalarmasses}
    \eeq
    
    The fields $h'$ and $s'$ are two real scalars and $\sigma_\phi$ and
    $\sigma_\chi$ are the corresponding Goldstone bosons that are weak
    eigenstates.  We shall denote the mass eigenstates with $h$, $s$ and $\sigma_Z$,
    $\sigma_{Z'}$. These different eigenstates are related by the rotations
    \beq
    \binom{h}{s} = \textbf{Z}_S
    \binom{h'}{s'}
    \equiv \begin{pmatrix}
        \cos \tS & -\sin \tS
        \\
        \sin \tS & ~~ \cos \tS
    \end{pmatrix}
    \binom{h'}{s'}
    \eeq
    and
    \beq
    \binom{\sigma_Z}{\sigma_{Z'}} = \textbf{Z}_{\rm G}
    \binom{\sigma_\phi}{\sigma_\chi}
    \equiv \begin{pmatrix}
        \cos \tG & -\sin \tG
        \\
        \sin \tG & ~~ \cos \tG
    \end{pmatrix}
    \binom{\sigma_\phi}{\sigma_\chi}
    \eeq
    where $\tS$ and $\tG$ 
    are the scalar and Goldstone mixing angles that can
    be determined by the diagonalization of the mass matrix of the real scalars 
    and that of the neutral Goldstone bosons. 
    
    The scalar mixing angle $\tS$ is related to the potential parameters by \cite{Trocsanyi:2018bkm}
    \beq
    \tan(2\tS) = -\frac{\lambda v w}
    {\lambda_\phi v^2 - \lambda_\chi w^2}
    \,.
    \eeq
    The condition $\tS \in (-\frac\pi4,\frac\pi4)$ implies that the scalar mass 
    eigenstates are not labeled by mass hierarchy.
    
    The mass matrix of the Goldstone bosons is given in principle by the sum of
    gauge-independent and gauge-dependent terms. However, the gauge-independent
    terms vanish by  Eq.~\eqref{eq:scalarmasses}:
    \beq
    \left( \begin{array}{cc}
        \half \lambda w^{2} + \lambda_\phi v^{2} - \mu_\phi^2  &0\\ 
        0 &\half \lambda v^{2} + \lambda_\chi w^{2} - \mu_\chi^2
    \end{array} \right) = \textbf{0}\,,
    \eeq
    so the mass matrix contains only gauge-dependent terms,
    \beq 
    \textbf{m}^2_{A} = \xi_Z\textbf{m}^2_{A_Z} +
    \xi_{Z'}\textbf{m}^2_{A_{Z'}}
    \eeq
    where $\xi_Z$ and $\xi_{{Z'}}$ are the gauge parameters.
    The mass matrix is symmetric, so we can write it formally as
    \beq
    \textbf{m}^2_{A_x} = 
    \left( \begin{array}{cc}
        m^2_{A_x,11} & m^2_{A_x,12}
        \\
        m^2_{A_x,12} & m^2_{A_x,22}
    \end{array} \right),
    \eeq
    for both $x = Z$ and ${Z'}$. Explicitly,
    \beq
    \bsp
    m^2_{A_Z,11} &=
    v^2 e^2 \Big(C^L_{Z\nu\nu} - C^R_{Z\nu\nu} \Big)^2 =
    \left(\frac{M_W}{\cos\tW}\right)^2 (\cos\tT - \kappa \sin\tT)^2
    \,,\\ 
    m^2_{A_Z,12} &=
    2 v w e^2 \Big(C^L_{Z\nu\nu} - C^R_{Z\nu\nu} \Big) C^R_{Z\nu\nu} =
    \left(\frac{M_W}{\cos\tW}\right)^2 (\cos\tT - \kappa \sin\tT) (-\tau \sin\tT)
    \,,\\
    m^2_{A_Z,22} &=
    w^2 e^2 \Big(2 C^R_{Z\nu\nu} \Big)^2 =
    \left(\frac{M_W}{\cos\tW}\right)^2 (-\tau \sin\tT)^2
    \label{eq:mAij}
    \esp
    \eeq
    where $M_W = \frac{v \gL}2$ is the mass of the W bosons, and the
    elements of $\textbf{m}^2_{A_{Z'}}$ can be obtained by the replacement
    $Z\to Z'$ in the chiral couplings, which implies the replacement
    \eqref{eq:ZtoT}
    in the second forms of the matrix elements. The latter are the most convenient
    ones for the diagonalization of the mass matrix. Using Eq.~\eqref{eq:sT-cT},
    one can check that the matrix
    \[
    \textbf{Z}_{\rm G} \textbf{m}^2_{A} \textbf{Z}_{\rm G}^T =
    \textbf{m}^2_{\text{diag},A} 
    \]
    is indeed diagonal provided we have for the Goldstone mixing angle
    \beq
    \cos \tG = \frac{\cos\tT-\kappa \sin\tT}
    {\sqrt{(\cos\tT - \kappa \sin\tT)^2 + (\tau \sin\tT)^2}}
    \eeq
    and
    \beq
    \sin \tG = \frac{\tau \sin\tT}
    {\sqrt{(\cos\tT-\kappa \sin\tT)^2 + (\tau \sin\tT)^2}}
    \,.
    \eeq

    \subsection{Masses of neutral gauge bosons}
    \label{sec:ZTmasses}
    
    As expected, the elements of the diagonal matrix
    $\textbf{m}^2_{\text{diag},A}$ coincide with the squares of the masses
    of the neutral gauge bosons \cite{Trocsanyi:2018bkm},
    \beq
    M_{Z}^2 = \left(\frac{M_{W}}{\cos\tW}\right)^2
    \Big[ (\cos\tT - \kappa \sin\tT)^2 + (\tau\sin\tT)^2 \Big]
    \label{eq:MZ0}
    \eeq
    and
    \beq
    M_{Z'}^2 = \left(\frac{M_{W}}{\cos\tW}\right)^2
    \Big[ (\sin\tT + \kappa \cos\tT)^2 + (\tau \cos\tT)^2 \Big]
    \,,
    \label{eq:MT0}
    \eeq
    which can also be expressed conveniently with the chiral couplings
    and Goldstone mixing angle.
    First we note that using Eq.~\eqref{eq:MZ0}, we find the simple relation
    \beq
    \sin \tG = \tau \frac{\sin\tT}{\cos\tW} \frac{M_W}{M_Z}
    \label{eq:sG}
    \eeq
    between the Goldstone and neutral boson mixing angles, and
    also
    \beq
    \cos \tG = \tau \frac{\cos\tT}{\cos\tW} \frac{M_W}{M_{Z'}}
    \,.
    \label{eq:cG}
    \eeq
    Next, we can substitute the relations found in Eq.~\eqref{eq:mAij} into
    Eqs.~\eqref{eq:MZ0} and \eqref{eq:MT0} together with the definition of the right handed
    couplings defined in Eq.~\eqref{eq:eCLR}, resulting in
    \beq
    M_Z^2 = v^2 e^2 \Big(C^L_{Z\nu\nu} - C^R_{Z\nu\nu} \Big)^2 +
    w^2 g_z^{\prime\,2} \sin^2\tT
    \eeq
    and also using Eq.~\eqref{eq:ZtoT},
    \beq
    M_{Z'}^2 = v^2 e^2 \Big(C^L_{Z'\nu\nu} - C^R_{Z'\nu\nu} \Big)^2 +
    w^2 g_z^{\prime\,2} \cos^2\tT
    \,.
    \eeq
    From Eq.~\eqref{eq:sG} and \eqref{eq:cG} we can express
    \beq
    w g_z' \sin\tT = M_Z\sin\tG
    \textrm{~~and~~}
    w g_z' \cos\tT = M_{Z'}\cos\tG
    \,,
    \label{eq:tG-tT}
    \eeq
    which after substitution and simple rearrangement leads to
    \beq
    M_Z^2 = \frac{v^2 e^2}{\cos^2\tG} \Big(C^L_{Z\nu\nu} - C^R_{Z\nu\nu} \Big)^2
    \,,\quad
    M_{Z'}^2 = \frac{v^2 e^2}{\sin^2\tG} \Big(C^L_{Z'\nu\nu} - C^R_{Z'\nu\nu} \Big)^2
    \,.
    \label{eq:MZ-cG}
    \eeq
    
    \subsection{Mass terms and mixing of neutrinos}
    \label{sec:numasses}
    
    The masses of the neutrinos are generated by the leptonic Yukawa terms in
    the Lagrangian \cite{Trocsanyi:2018bkm},
    \beq
    -\cL_Y^\ell = \half \overline{\nu_{R}^{c}}\:\textbf{Y}_N\:\nu_{R} \chi
    + \overline{L_{L}}\:\phi^c \:\textbf{Y}_\nu\:\nu_{R} + \hc 
    \eeq
    where $\overline{L_{L}}$ is the Dirac adjoint of the left handed lepton
    dublet, $\textbf{Y}_N$ and $\textbf{Y}_\nu$ are $3\times 3$ matrices, 
    the superscript $c$ denotes charge conjugation,
    $\nu^c = -\ri \gamma_2 \nu^*$.  After SSB this Lagrangian becomes
    \beq
    -\cL_Y^\ell = \frac{w+s'+\ri\sigma_\chi}{2\sqrt{2}}
    \overline{\nu_{R}^{c}} \:\textbf{Y}_N\: \nu_{R}
    + \frac{v+h'-\ri\sigma_\phi}{\sqrt{2}}
    \overline{\nu_{L}} \:\textbf{Y}_\nu\: \nu_{R}
    + \hc
    \label{eq:nuYukawa}
    \eeq
    and the terms proportional to the VEVs provide the mass matrices
    \beq
    \textbf{M}_N = \frac{w}{\sqrt{2}}\textbf{Y}_N
    \,,\quad
    \textbf{M}_D = \frac{v}{\sqrt{2}}\textbf{Y}_\nu
    \eeq
    where the Majorana mass matrix $\textbf{M}_N$ is real and symmetric, while
    the Dirac mass matrix $\textbf{M}_D$ is complex and Hermitian.
    
    In flavour basis the $6\times 6$ mass matrix for the neutrinos that can be
    written in terms of $3\times 3$ blocks as
    \beq
    \textbf{M}' =
    \begin{pmatrix}
        \textbf{0}_3 & \textbf{M}_D^T \\
        \textbf{M}_D & \textbf{M}_N
    \end{pmatrix}.
    \label{eq:Mflavour}
    \eeq
    The weak (flavour) eigenstates
    $(\nu_e,\,\nu_\mu,\,\nu_\tau,\,\nu_{R,1},\,\nu_{R,2},\,\nu_{R,3})$
    can be transformed into the basis of $\nu_i$ ($i=1-6$)
    mass eigenstates with a $6\times 6$ unitary matrix \footnote{This matrix
        coincides with the unitary matrix $U^{V,\dagger}$ used by \SARAH.}
    $\textbf{U}$ where the mass matrix is diagonal,
    \beq
    \textbf{U}^T\textbf{M}'\textbf{U} = \textbf{M}
    = \operatorname{diag}(m_1,m_2,m_3,m_4,m_5,m_6)
    \,.
    \label{eq:diag}
    \eeq
    It is helpful to decompose the matrix $\textbf{U}$ into two $3\times 6$
    blocks $\textbf{U}_L$ and $\textbf{U}_R^*$,
    \beq
    \textbf{U} = \binom{\textbf{U}_L}{\textbf{U}_R^*}
    \,,
    \label{eq:U=ULUR*}
    \eeq
    so $\textbf{U}^T = (\textbf{U}_L^T, \textbf{U}_R^\dag)$ where both blocks
    are $6\times 3$ matrices. It may be worth to emphasize that in spite
    of what might be implied by the notation, the matrices $\textbf{U}_L$
    and $\textbf{U}_R^*$ are only semi-unitary. Useful relations of these
    matrices are collected in \app{app:UMrelations}.
    
    \subsection{Gauge boson -- neutrino interactions}
    \label{sec:ZTnuinteractions}
    
    As the neutral currents are written in terms of flavour eigenstates,
    the interactions between the neutral gauge bosons and the propagating
    mass eigenstate neutrinos include also the neutrino mixing matrices:
    \beq
    \bom{\Gamma}^\mu_{V\nub_i\nu_j} = 
    -\ri e \gamma^\mu
    \Big(\bom{\Gamma}^L_{V\nub\nu} P_L + \bom{\Gamma}^R_{V\nub\nu} P_R\Big)_{ij}
    \label{eq:GVnn}
    \eeq
    where
    \beq
    \bom{\Gamma}^L_{V\nub\nu} = 
    C^L_{V\nu\nu}\textbf{U}_L^\dagger \textbf{U}_L
    -C^R_{V\nu\nu}\textbf{U}_R^T\textbf{U}_R^*
    \label{eq:GLVnn}
    \eeq
    and 
    \beq\label{eq:GRVnn}
    \bom{\Gamma}^R_{V\nub\nu} = 
    -C^L_{V\nu\nu}\textbf{U}_L^T \textbf{U}^*_L
    +C^R_{V\nu\nu}\textbf{U}_R^\dagger\textbf{U}_R =
    - \Big(\bom{\Gamma}^L_{V\nub\nu}\Big)^*
    \eeq
    for both $V=Z$ and $V=Z'$.
    
    \subsection{Scalar boson -- neutrino and Goldstone boson -- neutrino interactions}
    \label{sec:Snuinteractions}
    
    The terms containing the scalar and Goldstone bosons in Eq.~\eqref{eq:nuYukawa} provide 
    interactions between those and the neutrinos. These interactions have the same
    structure with small differences. For the propagating scalar states $S_k$ or
    $\sigma_k$ ($k=1$ denoting $h$ or the Goldstone boson belonging to $Z$ and $k=2$
    referring to $s$ or the Goldstone boson belonging to $Z'$) such interactions can 
    be decomposed into left and right chiral terms
    \beq
    \bom{\Gamma}_{S_k/\sigma_k\, \nu_i \nu_j} =
    \Big(\bom{\Gamma}_{S_k/\sigma_k\, \nu \nu}^L P_L +
    \bom{\Gamma}_{S_k/\sigma_k\, \nu \nu}^R P_R\Big)_{ij}
    \label{eq:GSnn}
    \eeq
    where the matrices $\bom{\Gamma}^{L/R}$ contain both the
    mixing matrix of the neutrinos and the mixing matrix of the scalar or Goldstone
    bosons. The left chiral coefficients are
    \begin{align} 
        \bom{\Gamma}^L_{S_k\nub\nu} &=
        -\ri\left[\Big(\textbf{M}\textbf{U}_L^\dagger\textbf{U}_L
        +\textbf{U}_L^T \textbf{U}_L^*\textbf{M}\Big) \frac{(\textbf{Z}_S)_{k1}}{v}
        + \textbf{U}_R^\dagger \textbf{M}_N \textbf{U}_R^*
        \frac{(\textbf{Z}_S)_{k2}}{w}\right]
        \,,
        \label{eq:GSLnn}
    \end{align}
    and 
    \begin{align} 
        \bom{\Gamma}^L_{\sigma_k\nub\nu} &=
        -\left[\Big(\textbf{M}\textbf{U}_L^\dagger\textbf{U}_L
        +\textbf{U}_L^T \textbf{U}_L^*\textbf{M}\Big) \frac{(\textbf{Z}_{\rm G})_{k1}}{v}
        + \textbf{U}_R^\dagger \textbf{M}_N \textbf{U}_R^*
        \frac{(\textbf{Z}_{\rm G})_{k2}}{w}\right]
    \end{align}
    and the right chiral ones are related by complex conjugation,
    $\bom{\Gamma}^R_{S_k/\sigma_k\, \nu \nu} =
    -\Big(\bom{\Gamma}^{L}_{S_k/\sigma_k\, \nu \nu}\Big)^*$.
    
    \section{Neutrino mass matrix at one-loop order}
    \label{sec:1loop}
    
    We are interested in the one-loop correction $\delta \textbf{M}_L$ to
    the tree-level mass matrix of the light neutrinos.
    In perturbation theory we deal with propagating states which are mass
    eigenstates. Hence, we can compute loop corrections to self energies
    of mass eigenstates of neutrinos. The neutrino mass matrix at one-loop
    order is then obtained from Eq.~\eqref{eq:diag}, with diagonal mass matrix
    substituted at one loop, $\textbf{M} + \delta \textbf{M}$ where 
    \beq
    \delta \textbf{M}
    = \operatorname{diag}(\delta m_1,\delta m_2,\delta m_3,\delta
    m_4,\delta m_5,\delta m_6)
    \,.
    \eeq
    Hence, the correction is obtained by
    \beq
    \delta\textbf{M}' =
    \begin{pmatrix}
        \delta \textbf{M}_L & \delta \textbf{M}_D^T \\
        \delta \textbf{M}_D & \delta \textbf{M}_N
    \end{pmatrix} = 
    \textbf{U}^*\delta \textbf{M}\textbf{U}^\dagger
    \,.
    \eeq
    Using Eq.~\eqref{eq:U=ULUR*}, we can compute the $3\times 3$ blocks as
    \beq
    \delta \textbf{M}_L = \textbf{U}_L^*\delta\textbf{M}\textbf{U}_L^\dagger
    ,\quad
    \delta \textbf{M}_D = \textbf{U}_R  \delta\textbf{M}\textbf{U}_L^\dagger
    ,\quad
    \delta \textbf{M}_N = \textbf{U}_R  \delta\textbf{M}\textbf{U}_R^T
    \,.
    \eeq 
    In the following subsections we prove that the one-loop correction to the 
    mass matrix of the active neutrinos have the form
    \beq
    \delta \textbf{M}_L = \frac{1}{16\pi^2}\sum_{k=1,2} 
    \left[
    3 (\textbf{Z}_{\rm G})_{k1}^2 \frac{M^2_{V_k}}{v^2} \textbf{F}(M^2_{V_k})
    + (\textbf{Z}_{\rm S})_{k1}^2 \frac{M^2_{S_k}}{v^2} \textbf{F}(M^2_{S_k})
    \right]
    \label{eq:finite-correction}
    \eeq
    where we introduced the finite matrix valued function
    \be 
    \textbf{F}_{ij}(M^2) = \sum_{a=1}^6 (\textbf{U}_L^*)_{ia}(\textbf{U}_L^\dagger)_{aj} 
    \frac{m_a^3}{M^2}\frac{\ln \frac{m_a^2}{M^2}}{\frac{m_a^2}{M^2}-1}
    \label{eq:fmatrixfunction}
    \ee 
    of dimension mass and with summation running over all neutrinos.
    
    \subsection{Self-energy decomposition}
    
    The neutrino self energy is a $6\times 6$ matrix that can be
    decomposed as 
    \beq
    \ri \bom{\Sigma}(p) =
    \textbf{A}_L(p^2) \slashed{p} P_L + \textbf{A}_R(p^2) \slashed{p} P_R
    + \textbf{B}_L(p^2) P_L + \textbf{B}_R(p^2) P_R
    \,.
    \eeq
    Using this decomposition, $\delta \textbf{M}_L$ is given by \cite{Grimus:2002nk}
    \beq\label{eq:loopmasscorr}
    \delta \textbf{M}_L =
    \textbf{U}_L^*\textbf{B}_L(0)\textbf{U}_L^\dagger
    \,.
    \eeq
    
\begin{figure}[th]
        \begin{center}
            \includegraphics[width=0.4\linewidth]{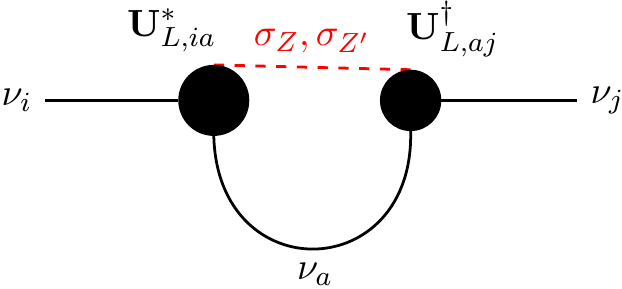}
            \includegraphics[width=0.4\linewidth]{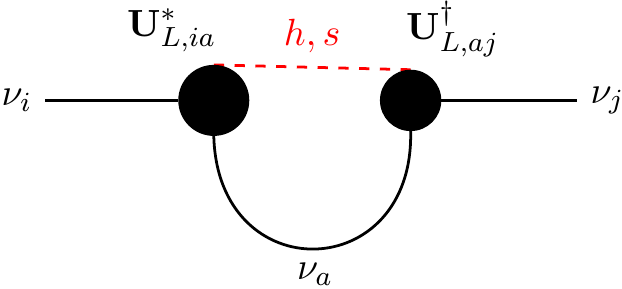}
            
            \includegraphics[width=0.4\linewidth]{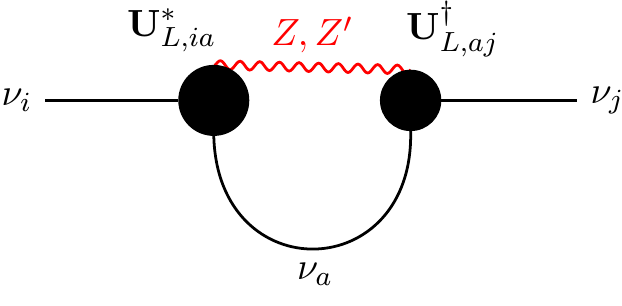}
            \includegraphics[width=0.4\linewidth]{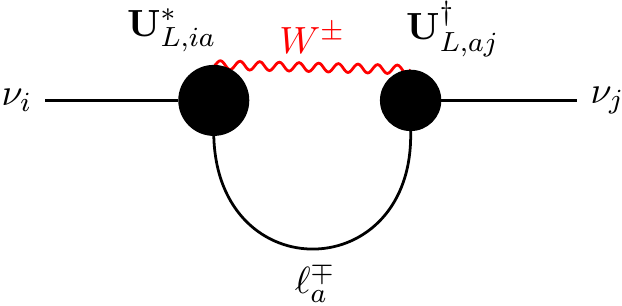}
        \end{center}
        \caption{\label{fig:loop}
            One-loop graphs contributing to the neutrino self energy. 
            Top left: Goldstone boson contribution. Top right: scalar contribution. 
            Bottom left: neutral gauge boson contribution. Bottom right: charged gauge 
            boson contribution. Note that $W$ boson loop does not contribute to the 
            matrix \textbf{B}$_L$.}
    \end{figure}

    The matrix $\textbf{B}_L(0)$ receives contributions involving
    a neutrino and either a neutral vector boson Z, Z', or a scalar boson
    $\sigma_Z$, $\sigma_{Z'}$ (Goldstone boson), $h$, $s$ (Higgs-like scalar) in the
    loop.  The relevant Feynman graphs that give contributions to the
    neutrino self energies at one-loop order are shown in \fig{fig:loop}.
    There are also tadpole contributions to $\textbf{B}_L(0)$. 
    Those are proportional to the scalar-neutrino coupling
    $\bom{\Gamma}^L_{S_k\nub_i\nu_j}$ given in Eq.~\eqref{eq:GSnn}, 
    which vanishes when sandwiched between $\textbf{U}_L^*$ and
    $\textbf{U}_L^\dagger$, see Eq.~\eqref{eq:UL*MULd}. The charged 
    vector boson together with a charged lepton in the loop (bottom right 
    diagram in \fig{fig:loop}) contributes only to $\textbf{A}_{L/R}$. 
    Thus we compute the first three graphs explicitly.
    For a given boson $x$ in the loop, the matrix $\textbf{B}_L(0)$ depends
    on the mass $M_x$ and also the tree-level masses of the neutrinos
    $\{m_a\}$, $\textbf{B}_L(0) = \textbf{B}_L^x(M_x,\{m_a\})$.
    
    \subsection{Contributions with neutral gauge bosons in the loop}
    \label{sec:Vcont}
    
    The contribution of the neutral gauge boson $V$ is
    \beq
    \Big(B_L^V(M_V,\{m_a\};\xi_V)\Big)_{ij} P_L =
    \ri\int\!\frac{\rd^d\ell}{(2\pi)^d}
    \sum_{a=1}^6
    \Gamma^{\mu}_{V\nub_i\nu_a}
    \frac{\slashed{p}-\slashed{\ell}+m_a}{(p-\ell)^2-m_a^2}
    \Gamma^\nu_{V\nu_a\nub_j} P_{\mu\nu}(\ell,M_V^2;\xi_V)
    \eeq
    where $\xi_V$ is the gauge parameter and
    \beq
    P_{\mu\nu}(\ell,M_V^2;\xi_V) = 
    \frac{g_{\mu\nu}}{\ell^2-M_V^2} - (1-\xi_V)
    \frac{\ell_\mu \ell_\nu}{(\ell^2-M_V^2)(\ell^2-\xi_V M_V^2)}
    \,.
    \eeq 
    Introducing the $6\times 6$ matrix 
    \beq
    \textbf{m}_\ell^{(n)} = \operatorname{diag}
    \left( \frac{m_1^n}{\ell^2-m_1^2}, \dots, \frac{m_6^n}{\ell^2-m_6^2}\right),
    \eeq
    and using the result of \app{app:proof}, we obtain the following 
    expression for a neutral vector boson in the loop:
    \beq
    \label{eq:vectorcontribution}
    \delta \textbf{M}_L^V = 
    \ri e^2 \Big(C^L_{V\nu\nu}-C^R_{V\nu\nu}\Big)^2 \int\!\frac{\rd^d\ell}{(2\pi)^d}
    \textbf{U}_L^*
    \bigg[
    \frac{d\:\textbf{m}_\ell^{(1)}}{\ell^2-M_V^2}
    + \frac{\textbf{m}_\ell^{(3)}}{M_V^2} \left(
    \frac{1}{\ell^2-\xi_V M_V^2} - \frac{1}{\ell^2-M_V^2}\right)\bigg]
    \textbf{U}_L^\dagger
    .
    \eeq
    
    \subsection{Contributions with neutral Goldstone bosons in the loop}
    \label{sec:Gcont}
    
    The contribution of the neutral Goldstone boson $\sigma_V$ ($V=1$ means the 
    Goldstone boson belonging to the $Z$ field and $V=2$ refers to the $Z'$ field) is 
    \beq
    \Big(B_L^{\sigma_V}(m_{\sigma_V},\{m_a\};\xi_V)\Big)_{ij} P_L =
    - \ri\int\!\frac{\rd^d\ell}{(2\pi)^d}
    \sum_{a=1}^6
    \bom{\Gamma}_{\sigma_V\nub_i\nu_a}
    \frac{m_a}{\ell^2-m_a^2}
    \bom{\Gamma}_{\sigma_V\nu_a\nub_j} \frac{1}{\ell^2 - \xi_V M^2_V}
    \,.
    \eeq
    Using the matrix notation, we can write
    \beq
    \textbf{U}_L^*\textbf{B}_L^{\sigma_V}(0)\textbf{U}_L^\dagger P_L = 
    - \ri \int\!\frac{\rd^d\ell}{(2\pi)^d}
    \textbf{U}_L^*
    \bom{\Gamma}_{\sigma_V\nub\nu}
    \textbf{m}_\ell^{(1)}
    \bom{\Gamma}_{\sigma_V\nu\nub}
    \textbf{U}_L^\dagger
    \frac{1}{\ell^2 - \xi_V M^2_V}
    \,.
    \eeq
    Substituting the vertex functions of Eq.~\eqref{eq:GSnn} and employing the
    matrix relations in Eqs.~\eqref{eq:Urel1} and \eqref{eq:UL*MULd}, we obtain the
    correction to the mass matrix as
    \beq
    \delta \textbf{M}_L^{\sigma_V} = 
    - \ri \int\!\frac{\rd^d\ell}{(2\pi)^d}
    \textbf{U}_L^*
    \textbf{M}
    \textbf{m}_\ell^{(1)}
    \textbf{M}
    \textbf{U}_L^\dagger
    \left(\frac{(\textbf{Z}_{\rm G})_{V1}}{v}\right)^2
    \frac{1}{\ell^2 - \xi_V M^2_V}
    \,.
    \label{eq:deltaMLsigma}
    \eeq
    We now substitute
    $\textbf{M} \textbf{m}_\ell^{(1)} \textbf{M} = \textbf{m}_\ell^{(3)}$
    and using Eq.~\eqref{eq:MZ-cG},
    we obtain
    \beq
    \label{eq:goldstonecontribution}
    \delta \textbf{M}_L^{\sigma_V} = 
    - \ri e^2
    \Big(C^L_{V\nu\nu} - C^R_{V\nu\nu} \Big)^2
    \int\!\frac{\rd^d\ell}{(2\pi)^d}
    \textbf{U}_L^*
    \frac{\textbf{m}_\ell^{(3)}}{M_V^2}
    \textbf{U}_L^\dagger
    \frac{1}{\ell^2 - \xi_V M^2_V}
    \,.
    \eeq
    
    \subsection{Contributions with scalar bosons in the loop}
    \label{sec:Scont}
    
    The scalar -- neutrino vertex is very similar to the Goldstone boson neutrino
    vertex, so the contribution with a scalar boson $S_k$ in the loop can be written
    immediately in analogy with Eq.~\eqref{eq:deltaMLsigma}:
    \beq
    \bsp
    \label{eq:scalarcontribution}
    \delta \textbf{M}_L^{S_k} &= 
    \ri \int\!\frac{\rd^d\ell}{(2\pi)^d}
    \textbf{U}_L^*
    \textbf{M}
    \textbf{m}_\ell^{(1)}
    \textbf{M}
    \textbf{U}_L^\dagger
    \left(\frac{(\textbf{Z}_S)_{k1}}{v}\right)^2
    \frac{1}{\ell^2 - M^2_{S_k}}
    \\ &=
    \ri \left(\frac{(\textbf{Z}_S)_{k1}}{v}\right)^2
    \int\!\frac{\rd^d\ell}{(2\pi)^d}
    \textbf{U}_L^*
    \textbf{m}_\ell^{(3)}
    \textbf{U}_L^\dagger
    \frac{1}{\ell^2 - M^2_{S_k}}
    \,.
    \esp
    \eeq
    
    \subsection{The complete one-loop mass correction}
    
    Combining Eqs.~\eqref{eq:vectorcontribution}, \eqref{eq:goldstonecontribution} 
    and \eqref{eq:scalarcontribution}, we find that that the gauge-dependent pieces 
    of the vector boson contribution cancel exactly with the Goldstone boson
    contribution, and obtain
    \beq
    \bsp
    \delta \textbf{M}_L &= 
    \sum \limits_{V=Z,Z'}\brac{\delta \textbf{M}_L^V + \delta \textbf{M}_L^{\sigma_V}}
    + \sum \limits_{k=1,2}\delta \textbf{M}_L^{S_k}
    \\&=
    \sum \limits_{V=Z,Z'}\ri e^2 \Big(C^L_{V\nu\nu}-C^R_{V\nu\nu}\Big)^2 
    \int\!\frac{\rd^d\ell}{(2\pi)^d}
    \textbf{U}_L^*
    \bigg[
    \frac{d\:\textbf{m}_\ell^{(1)}}{\ell^2-M_V^2}
    - \frac{\textbf{m}_\ell^{(3)}}{M_V^2} 
    \frac{1}{\ell^2-M_V^2}\bigg]
    \textbf{U}_L^\dagger
    \\&\quad+
    \sum \limits_{k=1,2}
    \ri \left(\frac{(\textbf{Z}_S)_{k1}}{v}\right)^2
    \int\!\frac{\rd^d\ell}{(2\pi)^d}
    \textbf{U}_L^*
    \frac{\textbf{m}_\ell^{(3)}}{\ell^2 - M^2_{S_k}}
    \textbf{U}_L^\dagger
    \,.
    \label{eq:oneloopselfenergy}
    \esp
    \eeq 
    Introducing the integral
    \beq
    \label{eq:I0-first}
    I_0(M^2,m_a^2;\mu^2,\epsilon) = \mu^{2\epsilon}
    \int \frac{\rd^d\ell}{(2\pi)^d}\frac{1}{(\ell^2-M^2)(\ell^2-m_a^2)}\,,
    \eeq
    the matrix $\textbf{I}^{(n)}$ with elements
    \bal 
    \label{eq:integralmatrix}
    \textbf{I}^{(n)}_{ij}(M^2) = 
    \ri\sum\limits_{a=1}^6 (\textbf{U}_L^*)_{ia}m_a^n(\textbf{U}_L^\dagger)_{aj} 
    I_0(M^2,m_a^2;\mu^2,\epsilon)\,,
    \eal 
    and using the relations \eqref{eq:MZ-cG} allows us to recast
    Eq.~\eqref{eq:oneloopselfenergy} into a neatly condensed form
    \bal
    \delta \textbf{M}_L 
    &= 
    \sum \limits_{k=1,2}\left\{\left(\frac{(\textbf{Z}_{\rm G})_{k1}}{v}\right)^2\brac{d M_{V_k}^2 \textbf{I}^{(1)}(M_{V_k}^2) - \textbf{I}^{(3)}(M_{V_k}^2)}
    +\brac{\frac{(\textbf{Z}_S)_{k1}}{v}}^2 \textbf{I}^{(3)}(M_{S_k}^2)
    \right\}
    \eal  
    with $V_1 = Z$ and $V_2= Z'$. In Eq.~\eqref{eq:I0-first} $2\epsilon = d-4$ 
    and $\mu$ is the regularization scale.
    
    \subsection[Finiteness and scale independence of delta-ML]{Finiteness and scale independence of $\delta\textbf{M}_L$}
    
    We show here that the one loop mass correction $\delta\textbf{M}_L$ is finite 
    and independent of the scale $\mu$.
    Evaluating the integral \eqref{eq:I0-first} yields
    \bal \label{eq:I0-second}
    I_0(M^2,m^2;\mu^2,\epsilon) &= 
    I^{(\rs)}_0(\epsilon) + I^{(\rf)}_0\left(\frac{m^2}{M^2},\frac{\mu^2}{M^2}\right)
    +\rO(\epsilon)
    \eal
    where `s' stands for the singular and `f' for the finite functions 
    \bal
    I^{(\rs)}_0(\epsilon) =\frac{\ri}{16\pi^2}
    \brac{\frac{1}{\epsilon} - \gamma_\rE + \ln 4\pi + 1}
    &,\quad
    I^{(\rf)}_0(x,x_\mu) =\frac{\ri}{16\pi^2}
    \brac{\frac{x \ln x}{1-x} + \ln x_\mu}.
    \eal
    with $\gamma_\rE \simeq 0.5722$ being the Euler-Mascheroni constant. It is also
    convenient to split the matrix \eqref{eq:integralmatrix} in a similar fashion
    \beq
    \textbf{I}^{(n)}(M^2) =
    \textbf{I}^{(\rs,n)}+\textbf{I}^{(\rf,n)}(M^2)
    \eeq
    such that
    \beq
    \textbf{I}^{(\rs,n)} =
    \ri\textbf{U}_L^*\textbf{M}^n\textbf{U}_L^\dagger I^{(\rs,n)}_0  
    \,,\quad
    \Big(\textbf{I}^{(\rf,n)}(M^2)\Big)_{ij} = \ri\sum_{a=1}^6 (\textbf{U}_L^*)_{ia}m_a^n(\textbf{U}_L^\dagger)_{aj} 
    I^{(\rf,n)}_0\left(\frac{m_a^2}{M^2},\frac{\mu^2}{M^2}\right).
    \eeq
    Then the one-loop correction to the mass matrix of the light neutrinos can 
    also be decomposed as
    \beq
    \delta \textbf{M}_L = 
    \delta \textbf{M}^{(\rs)}_L 
    + \delta \textbf{M}^{(\rf)}_L + \rO(\epsilon)
    \eeq
    where 
    \bal
    \delta \textbf{M}^{(\rs)}_L &= 
    \sum_{k=1,2}\left[ d M_{V_k}^2\left(\frac{(\textbf{Z}_{\rm G})_{k1}}{v}\right)^2
    \textbf{I}^{(\rs,1)}
    +\frac{(\textbf{Z}_S)^2_{k1}-(\textbf{Z}_{\rm G})^2_{k1}}{v^2} \textbf{I}^{(\rs,3)}
    \right]
    \eal  
    and
    \bal
    \delta \textbf{M}^{(\rf)}_L 
    &= 
    \sum \limits_{k=1,2}\left[\left(\frac{(\textbf{Z}_{\rm G})_{k1}}{v}\right)^2
    \brac{d M_{V_k}^2 \textbf{I}^{(\rf,1)}(M_{V_k}^2) - \textbf{I}^{(\rf,3)}(M_{V_k}^2)}
    +\brac{\frac{(\textbf{Z}_S)_{k1}}{v}}^2 \textbf{I}^{(\rf,3)}(M_{S_k}^2)
    \right].
    \eal  
    In order to prove that $\delta\textbf{M}_L$ is finite, one has to show that 
    $\delta \textbf{M}^{(\rs)}_L$ is free from $\epsilon$ poles. We prove that 
    it in fact vanishes because the matrix $\textbf{I}^{(\rs,1)}$ is zero matrix 
    due to the identity \eqref{eq:UL*MULd}, while the coefficient in the second term
    cancels because the matrices \textbf{Z}$_S$ and \textbf{Z}$_G$ are orthogonal, so
    \beq
    \label{eq:mixing-matrix-cancellation}
    \sum_{k=1}^2 (\textbf{Z}_S)_{k1}^2 - \sum_{k=1}^2 (\textbf{Z}_G)_{k1}^2 = 0\,.
    \eeq 
    Hence the mass independent terms, including the divergent pieces of the light
    neutrino one-loop mass correction cancel, and we can set $\epsilon = 0$, which 
    yields $\delta\textbf{M}_L = \delta\textbf{M}^{(\rf)}_L$.
    Furthermore, the terms depending on the regularization scale in 
    $\delta \textbf{M}^{(\rf)}_L$ cancel in an identical way as the second term does 
    in $\delta \textbf{M}^{(\rs)}_L$ (using Eq.~\eqref{eq:mixing-matrix-cancellation}).
    
    The remaining finite terms give the final, regularization-scale independent and finite one-loop correction to the light neutrinos as given in Eq.~\eqref{eq:finite-correction}.
    It is the linear combination of the matrix valued function \textbf{F} 
    given in Eq.~\eqref{eq:fmatrixfunction} with different arguments and 
    coefficients corresponding to the different one-loop contributions.
    The function \textbf{F} gives the mass correction corresponding to a one 
    loop diagram, before coupling suppression; see Fig.~\ref{fig:loopmass} in 
    \sect{sec:numerics} for details where we shall also give a numerical estimate 
    for its eigenvalues $\delta m^\nu_{i,0}$. It is well defined for any non-negative
    $x$ because
    \beq
    \lim \limits_{x \rightarrow 0} \frac{x\ln x}{1-x} = 0
    \text{~~and~~}
    \lim \limits_{x \rightarrow 1} \frac{x\ln x}{1-x} = -1\,.
    \eeq
    
    \subsection{Generalization to arbitrary number of neutral bosons and neutrinos}
    
    Our predictions for the one-loop correction to the light neutrino mass matrix 
    can easily be generalized to any number of $n_V$ massive neutral gauge bosons, 
    $n_S$ neutral real scalars coupling to $n_a$ active and $n_s$ sterile neutrinos.
    Clearly, the matrix form of gauge-dependent parts in Eq.~\eqref{eq:vectorcontribution} and
    Eq.~\eqref{eq:goldstonecontribution} is unchanged, and they cancel in the same way.
    
    The correction without gauge parameters $\xi_V$ in Eq.~\eqref{eq:finite-correction} 
    is straightforwardly generalized to a case where the sums go over an arbitrary
    positive integer $n_V$ and $n_S$.
    
    The neutrino mass and mixing matrices with arbitrary $n_a$ and $n_s$ are written
    identically in the block form, differing only on the block shape: 
    \textbf{U}$_L$ is a $n_a \times (n_a+n_s)$ matrix and \textbf{U}$_R$ is a 
    $n_s \times (n_a+n_s)$ matrix. The finite correction derived in
    Eq.~\eqref{eq:finite-correction} is then immediately generalized to
    \beq
    \bsp
    \delta \textbf{M}_L = \frac{1}{16\pi^2}
    \left[
    3 \sum_{k=1}^{n_V} (\textbf{Z}_{\rm G})_{k1}^2 \frac{M^2_{V_k}}{v^2} \textbf{F}(M_{V_k}^2)
    + \sum_{k=1}^{n_S} (\textbf{Z}_{\rm S})_{k1}^2 \frac{M^2_{S_k}}{v^2} \textbf{F}(M_{S_k}^2)
    \right]
    \label{eq:generalized-finite-correction}
    \esp
    \eeq
    where the upper limit in the summation in the matrix $\textbf{F}$ is $n_a+n_s$.
    The factor 3 in front of the first term in the bracket of
    Eq.~\eqref{eq:generalized-finite-correction} stems from the three 
    polarization states of the propagating massive neutral gauge bosons. 
    The corresponding factor is of course one in the case of the scalars.
    This formula is also independent of the new U(1) charge assignments.

    \section{Numerical estimate of the corrections}
    \label{sec:numerics}
    
\begin{figure}[th]
        \centering
        \includegraphics[width=0.49\linewidth]{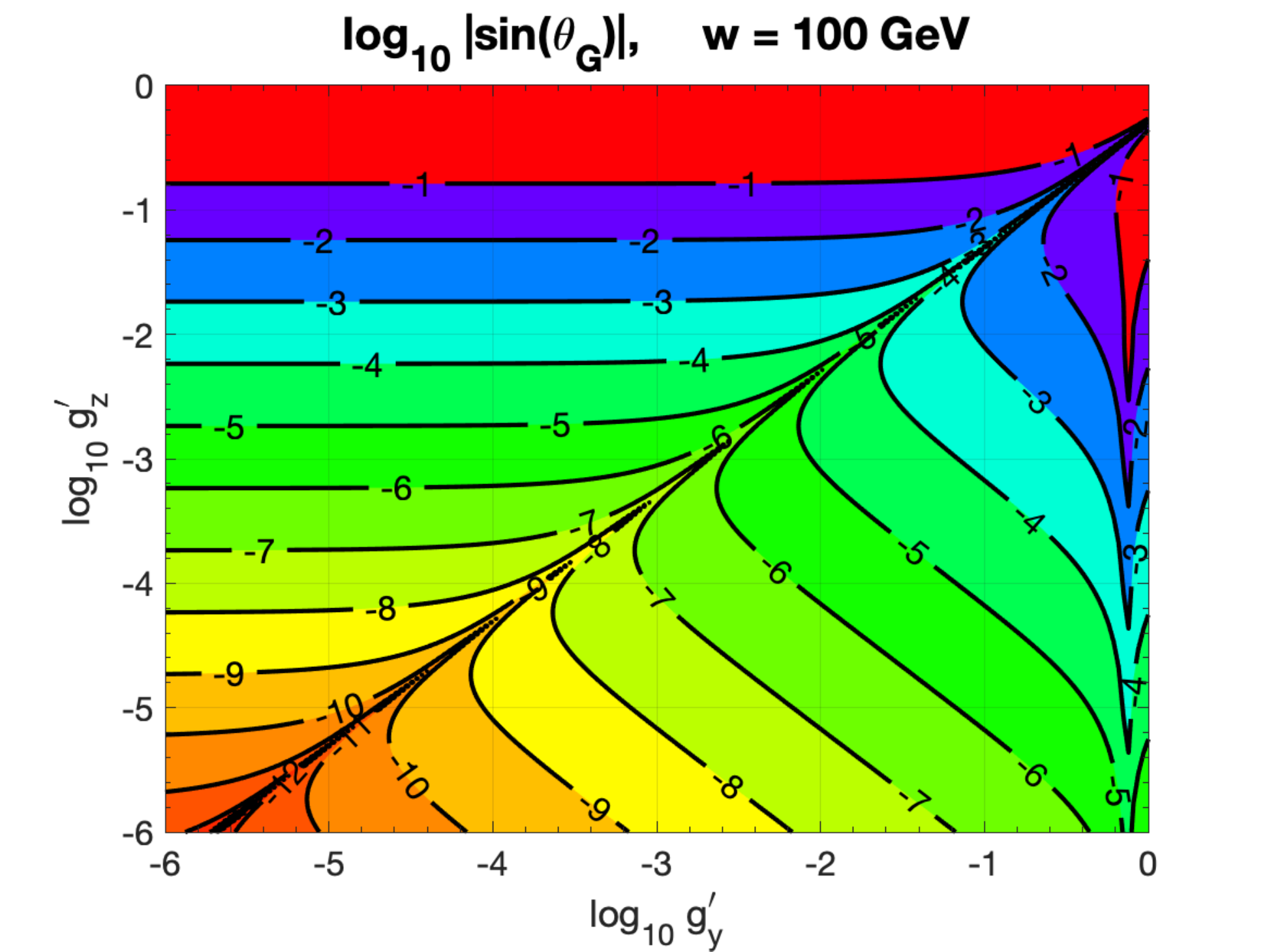}
        \includegraphics[width=0.49\linewidth]{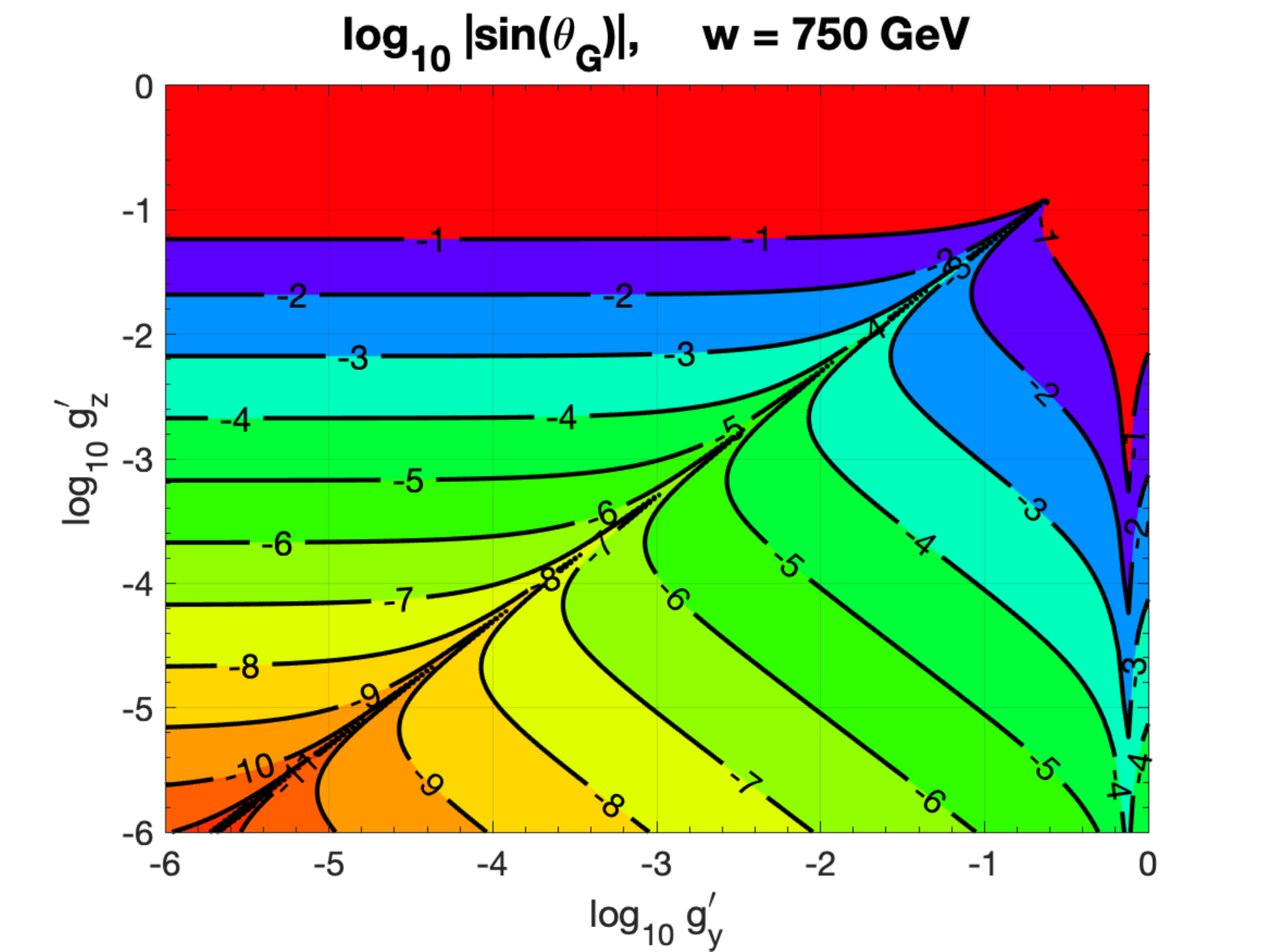}
        
        \includegraphics[width=0.49\linewidth]{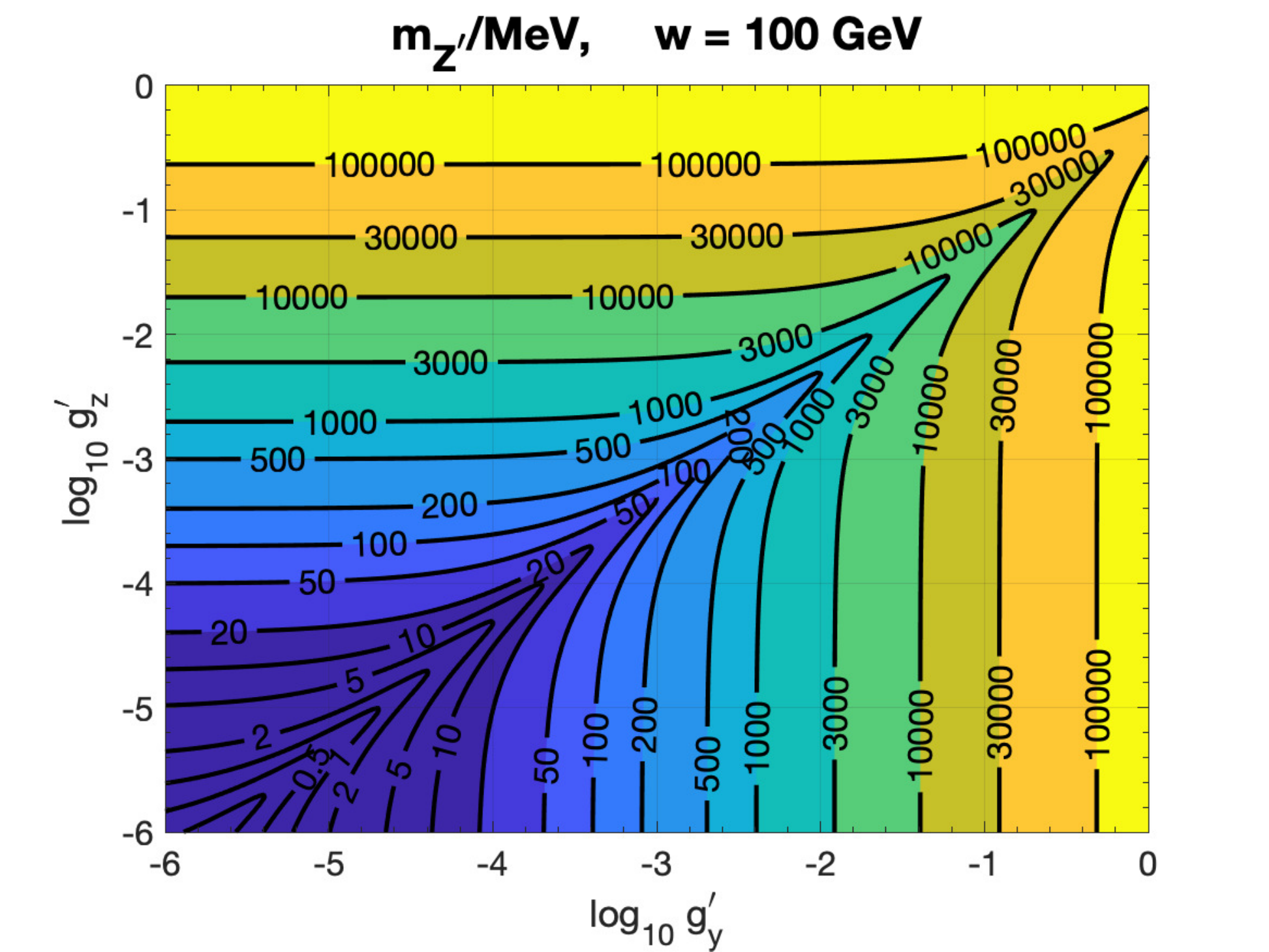} 
        \includegraphics[width=0.49\linewidth]{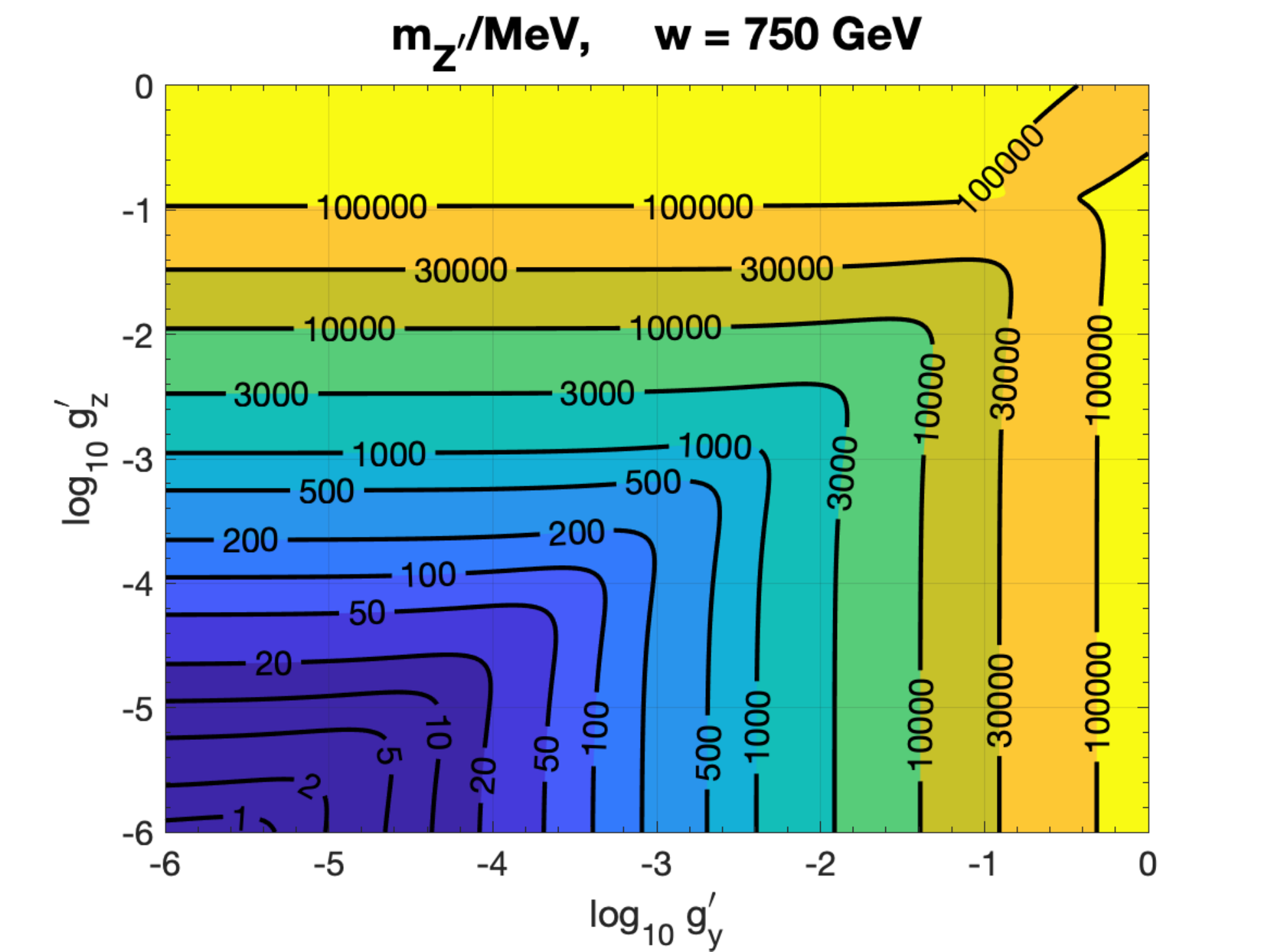} 
        \caption{\label{fig:contours} Absolute values of $\sin\theta_G$ (top) and 
            mass of $Z'$ boson (bottom) in logarithmic $(g_y',g_z')$ plane. 
            For $\theta_G$ the contour labels $n$ correspond to value $10^{n}$. 
            Left plots: $w = 100$\,GeV; right plots: $w= 750$\,GeV.
        }
    \end{figure}

    We are now ready to estimate the order of magnitude of the corrections.
    We assume large mixing in the scalar sector: $\theta_S = \rO(1)$.
    The $Z'$ mass and mixing angle $\theta_G$ are fixed by the gauge couplings 
    $g_y' = \gamma_y' g_L$ and $g_z'$ and ratio of VEVs, $\tan\beta \equiv w/v$.
    We plot their magnitudes in Fig. \ref{fig:contours}, scanning the parameters
    $g_y',g_z' \in [10^{-6},1]$ and for $w=100,750$ GeV. 
    Note that larger $\tan\beta$ corresponds to a larger Goldstone angle.  
    Smaller $\tan \beta$ distorts the $M_{Z'}$ contours so that the same $Z'$ mass 
    can be achieved with larger gauge couplings $g_y'$ and $g_z'$ compared to large 
    $\tan \beta$. In addition, we set $M_s/v = \rO(1)$, that is, 
    only the mass of the $Z'$ boson is free, and may be far from electroweak scale.
    The relevant gauge couplings can then be estimated as from Fig. \ref{fig:contours}
    after identifying the region in $(g_y',g_z')$ plane corresponding to 
    $M_{Z'} \in [20,200]$\,MeV, which is the relevant mass region for the 
    super-weak model to reproduce the dark matter relic density, allowed by 
    experimental constraints \cite{Iwamoto:2021fup}.
    
\begin{figure}
        \centering
        \includegraphics[width=0.49\linewidth]{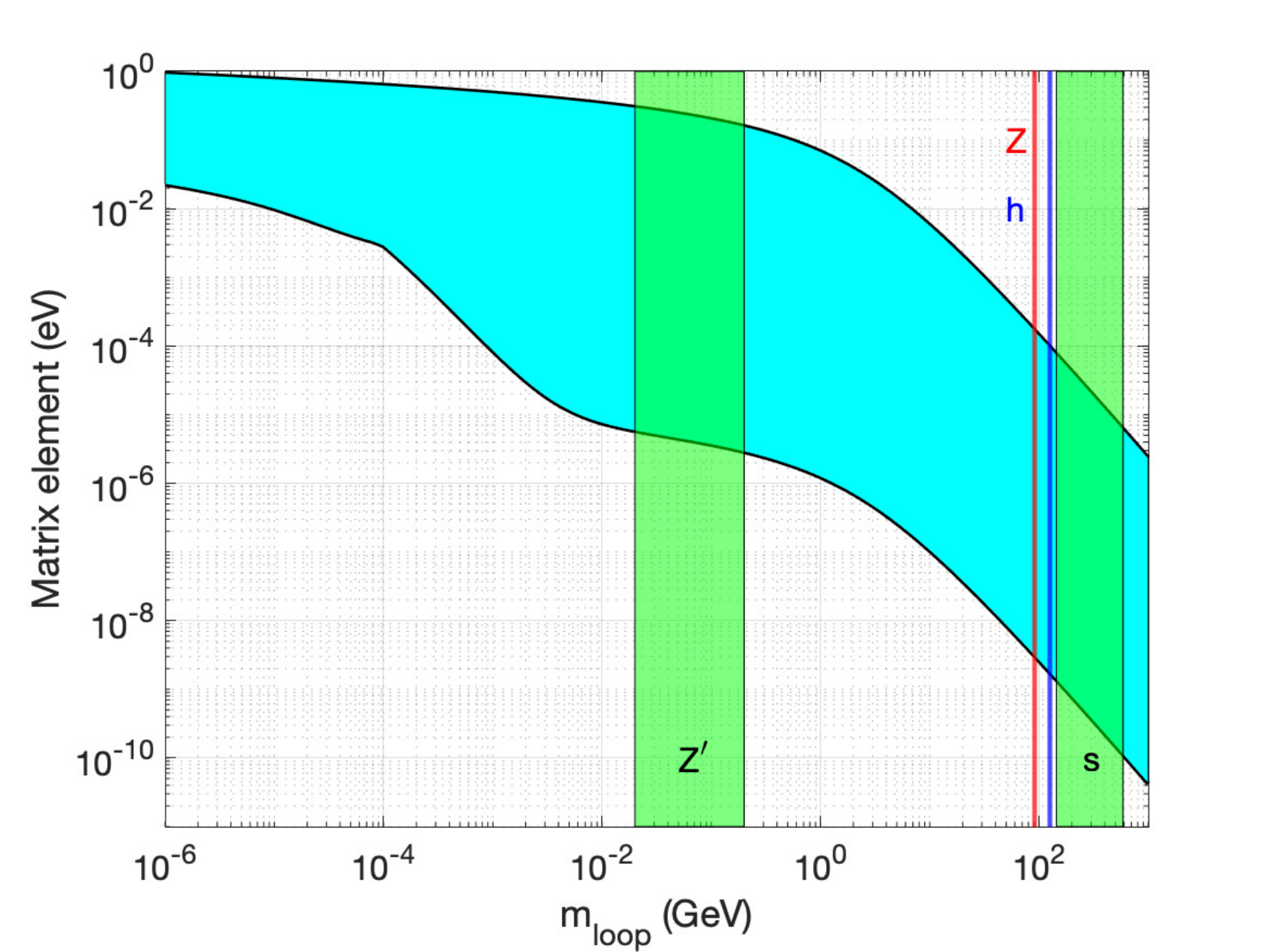}
        \includegraphics[width=0.49\linewidth]{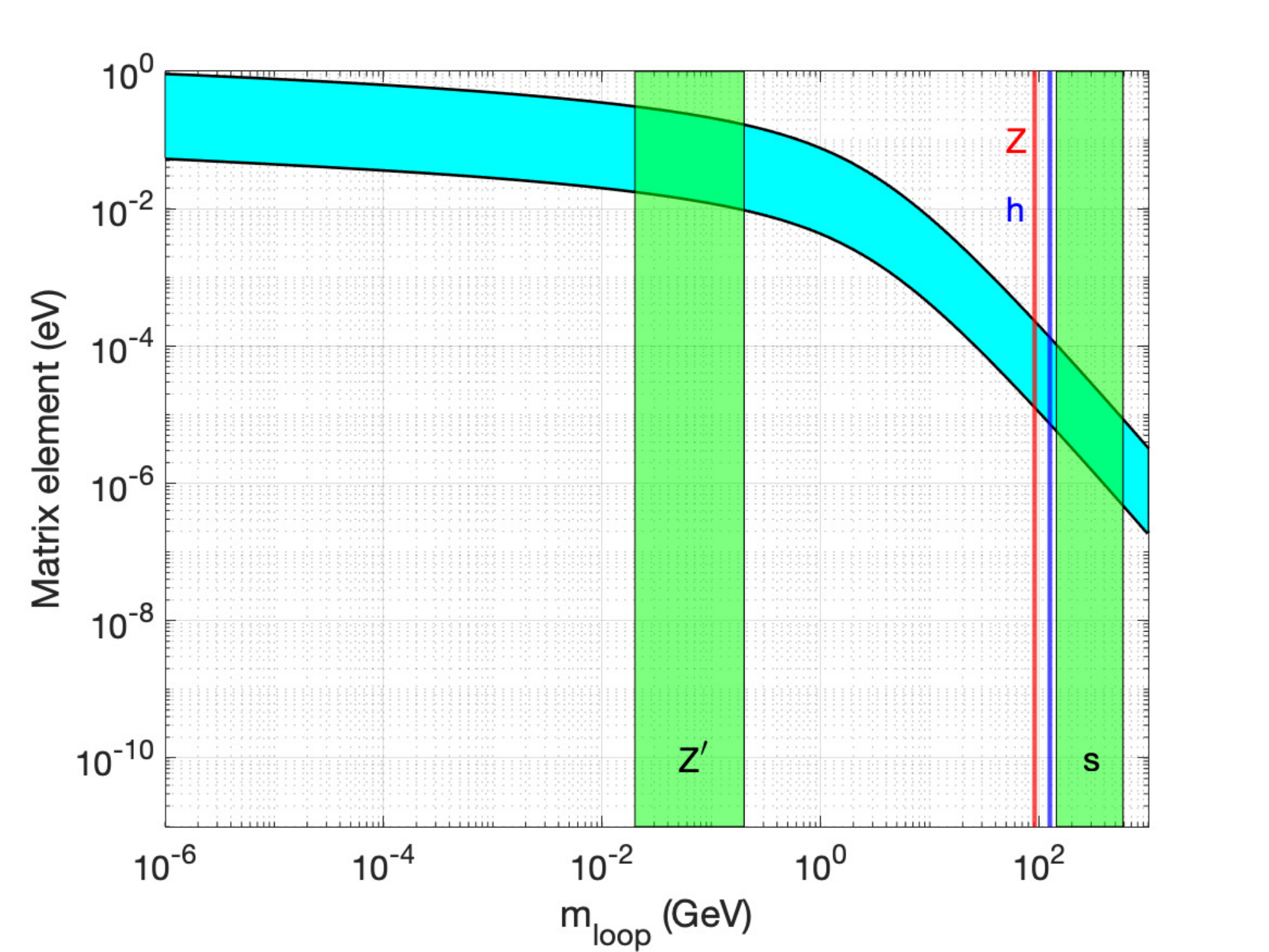}
        \caption{
            \label{fig:loopmatrix}
            Matrix elements \textbf{F}$_{ij}$ as a function of the mass $m_\text{loop}$ of 
            the boson in the loop are confined to the blue band, assuming normal neutrino mass
            hierarchy. We have highlighted with vertical bands the relevant mass regions where
            the masses of the bosons in the loop lie. The scalar $s$ is required to have
            mass between 144 and 558\,GeV requiring stability of the vacuum \cite{Peli:2019xwv}. 
            Left plot: $m_1^\text{tree} = 0.01$\,eV, $m_4^\text{tree} = 30$\,keV,
            $m_5^\text{tree} \approx m_6^\text{tree} = 2.5$\,GeV; 
            right plot: $m_1^\text{tree} = 0.001$\,eV, $m_4^\text{tree} = 7.1$\,keV,
            $m_5^\text{tree} \approx m_6^\text{tree} = 3.0$\,GeV.}
    \end{figure}

\begin{figure}
        \centering
        \includegraphics[width=0.8\linewidth]{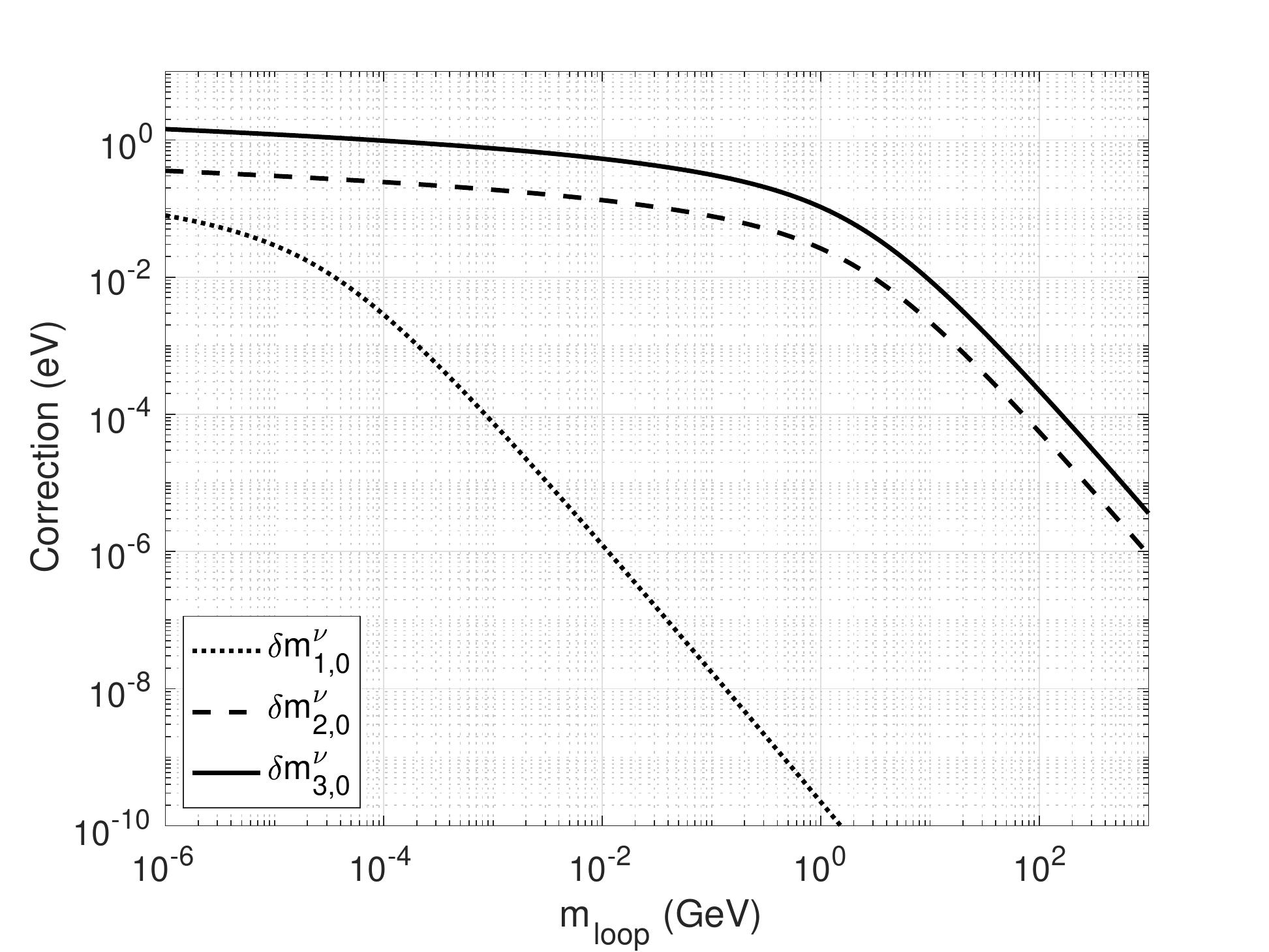}
        \caption{\label{fig:loopmass}
            Eigenvalues of the matrix \textbf{F} as a function of the mass of the boson 
            in the loop $m_\text{loop}$, assuming $m_1^\text{tree} = 0.01$ eV, 
            $m_4^\text{tree} = 30$ keV, $m_5^\text{tree} \approx m_6^\text{tree} = 2.5$ GeV, 
            and normal neutrino mass hierarchy.
        }
    \end{figure}

    Then we identify the order-of-magnitude estimate for 
    $|\sin\theta_G|$ by comparing the regions relevant to the mass range of $M_{Z'}$. 
    For $w=100$ GeV, we have $|\sin\theta_G| < 10^{-6}$, which we take as a
    conservative upper limit. Then the prefactors in gauge boson contributions to 
    $\delta \textbf{M}_L$ are
    
    \beq 
    e^2(C^L_{Z\nu\nu} - C^R_{Z\nu\nu})^2 = \cos^2\theta_G\frac{M_Z^2}{v^2} \sim \rO(10^{-1})
    \eeq
    and
    \beq
    e^2(C^L_{Z'\nu\nu} - C^R_{Z'\nu\nu})^2 = \sin^2\theta_G\frac{M_{Z'}^2}{v^2} \sim \rO(10^{-19})\times \brac{\frac{M_{Z'}}{100\,\text{MeV}}}^2.
    \eeq
    Then the numerical estimate for the total correction in Eq.~\eqref{eq:finite-correction} can be written as
    \begin{align}
        (\delta \textbf{M}_L)_{ij} 
        < \rO(10^{-7})\,\text{eV} + \rO(10^{-21})\times 
        \brac{\frac{M_{Z'}}{100\text{ MeV}}}^2 \textbf{F}_{ij}(M_{Z'}^2)
        \,.
        \label{eq:numeric}
    \end{align}
    The elements of the matrix \textbf{F} are plotted as a function of the mass 
    of the boson of the loop $m_{\rm loop}$ in
    Fig.~\ref{fig:loopmatrix} and the eigenvalues of the matrix corresponding to
    corrections to active neutrino species in Fig.~\ref{fig:loopmass}. 
    The eigenvalues of \textbf{F} themselves exceed the active neutrino tree-level
    masses, as the latter are at most about 10\,eV for the MeV scale $Z'$ boson. However, 
    the coupling suppressions in Eq.~\eqref{eq:numeric} are sufficient to tame the
    relative correction to the tree level mass below the per cent level.
    Assuming the active neutrino masses to be $\rO(10^{-3})$\,eV, a rough estimate 
    for the relative correction to active neutrino masses is of $\rO(10^{-4})$.
    
    We may maximize the effect of $Z'$ loop by allowing $Z'$ mass to be free and 
    setting large $|\sin \theta_G| = \rO(10^{-1})$, which is obtained when $g_y'$ and $g_z'$ are
    O($10^{-1}$). This corresponds to $M_{Z'} = \rO(M_Z)$, which is of course, excluded. 
    Yet, even in this case, the correction from $Z$ and $Z'$ loops are small, 
    have the same order of magnitude, $\rO(10^{-7})$\,eV. Thus, the individual
    contributions from BSM loops cannot be significantly larger than the SM
    contributions.

    \section{\label{sec:conclusions}Conclusions}
    
    In this paper, we have  computed the one-loop corrections to the mass matrix of 
    the active neutrinos in a gauged U(1) extension of the standard model of particle
    interactions. The field content of the model consists of a new complex scalar field 
    and three right-handed neutrinos---sterile under the standard model 
    interactions---in addition to the fields in the standard model. The neutrino
    masses are generated by Dirac and Majorana type Yukawa terms, which after
    spontaneous symmetry breaking of both scalar fields give rise to neutrino masses 
    in the way of the type I see-saw mass generation. We used $R_\xi$ gauge and have
    shown that the one-loop corrections are (i) independent of the gauge fixing
    parameters, (ii) finite and (iii) independent of the regularization scale.
    We also demonstrated how the formula for the one-loop mass corrections can be
    generalized to the case of arbitrary number of new U(1) groups, complex 
    scalars and right-handed neutrinos.
    
    We have provided a numerical estimate of the size of the mass corrections in 
    the context of the super-weak model, in which the new neutral gauge boson 
    $Z'$ is much lighter than the $Z$ boson of the standard model. We have 
    found that in the mass range of $M_{Z'} \in [20,200]$\,MeV, motivated by 
    a possible explanation of the relic density of dark matter in the Universe, 
    the relative mass corrections to the tree-level mass matrix elements do not 
    exceed the per mill level. Hence the model is stable against higher-order 
    corrections in the neutrino sector, which motivates further studies to 
    explore the viable parameter space of the model regarding the mixings 
    between the active and sterile neutrinos \cite{Karkkainen:2021}.
    
    \subsection*{Acknowledgments}
    
    We are grateful to Josu Hernández-García for discussions on this project.
    This work was supported by grant K 125105 of the National Research,
    Development and Innovation Fund in Hungary.
    \appendix
    \section{Some properties of neutrino mass and mixing matrices}
    \label{app:UMrelations}
    
    In this appendix, we derive some useful relations among the neutrino
    mass and mixing matrices. The matrix \textbf{U} that diagonalizes the
    neutrino mass matrix is unitary, hence
    \begin{equation}
        \textbf{UU}^\dagger =
        \binom{\textbf{U}_L}{\textbf{U}_R^*}(\textbf{U}_L^\dagger, \textbf{U}_R^T) =
        \begin{pmatrix}
            \textbf{U}_L \textbf{U}_L^\dagger & \textbf{U}_L \textbf{U}_R^T \\
            \textbf{U}_R^* \textbf{U}_L^\dagger & \textbf{U}_R^* \textbf{U}_R^T 
        \end{pmatrix} =
        \begin{pmatrix}
            \bom{1}_3 & \textbf{0}_3 \\ \textbf{0}_3 & \bom{1}_3
        \end{pmatrix},
    \end{equation}
    from which we obtain the following important relations:
    \beq
    \textbf{U}_L\textbf{U}_L^\dagger = \bom{1}_3
    \,,\quad
    \textbf{U}_R \textbf{U}_R^\dagger = \bom{1}_3
    \label{eq:Urel1}
    \eeq
    and
    \beq
    \textbf{U}_L\textbf{U}_R^T = \textbf{U}_R^*\textbf{U}_L^\dagger = \textbf{0}_3
    \label{eq:Urel2}
    \eeq 
    where $\bom{1}_n$ denotes the $n\times n$ unit matrix.
    The second unitarity conditions gives
    \beq
    \textbf{U}^\dagger \textbf{U} =
    (\textbf{U}_L^\dagger, \textbf{U}_R^T)\binom{\textbf{U}_L}{\textbf{U}_R^*} =
    \textbf{U}_L^\dagger \textbf{U}_L + \textbf{U}_R^T \textbf{U}_R^* =
    \bom{1}_6
    \,.
    \label{eq:Urel3}
    \eeq
    Using Eq.~\eqref{eq:diag} we derive
    \begin{align*}
        \textbf{U}_L^*\textbf{M}\textbf{U}_L^\dagger &=
        \textbf{U}_L^*(\textbf{U}_L^T, \textbf{U}_R^\dagger)\begin{pmatrix}
            \textbf{0}_3 & \textbf{M}_D^T \\ \textbf{M}_D & \textbf{M}_N
        \end{pmatrix}\binom{\textbf{U}_L}{\textbf{U}_R^*}\textbf{U}_L^\dagger\\
        &= \textbf{U}_L^*\textbf{U}_L^T \textbf{M}_D 
        \textbf{U}_R^* \textbf{U}_L^\dagger
        + 
        \textbf{U}_L^* \textbf{U}_R^\dagger
        \textbf{M}_D \textbf{U}_L \textbf{U}_L^\dagger
        + \textbf{U}_L^T \textbf{U}_R^\dagger \textbf{M}_N 
        \textbf{U}_R^* \textbf{U}_L^\dagger 
        \,,
    \end{align*}
    and then with relations in Eq.~\eqref{eq:Urel2} we obtain
    \beq
    \textbf{U}_L^*\textbf{M}\textbf{U}_L^\dagger = \bom{0}_3\,.
    \label{eq:UL*MULd}
    \eeq
    Analogous calculations yield
    \beq
    \textbf{U}_R\textbf{M}\textbf{U}_L^\dagger = \textbf{M}_D\,.
    \label{eq:URMULd}
    \eeq
    Multiplying Eq.~\eqref{eq:URMULd} with $\textbf{U}_R^\dagger$ from the
    left and using Eq.~\eqref{eq:Urel3}, we find
    \be 
    \textbf{U}_R^\dagger \textbf{M}_D =
    \textbf{U}_R^\dagger \textbf{U}_R\textbf{M}\textbf{U}_L^\dagger =
    (\bom{1}_6 - \textbf{U}_L^T\textbf{U}_L^*)\textbf{M}\textbf{U}_L^\dagger =
    \textbf{M}\textbf{U}_L^\dagger - \textbf{U}_L^T 
    \textbf{U}_L^*\textbf{M}\textbf{U}_L^\dagger 
    \ee
    where the last term vanishes by Eq.~\eqref{eq:UL*MULd}, so
    \beq
    \textbf{U}_R^\dagger \textbf{M}_D = \textbf{M}\textbf{U}_L^\dagger
    \,.
    \label{eq:URdMD=MULd}
    \eeq 
    Finally,
    \begin{align*}
        \textbf{U}_R\textbf{M}\textbf{U}_R^T &=
        \textbf{U}_R(\textbf{U}_L^T, \textbf{U}_R^\dagger)
        \begin{pmatrix}
            \textbf{0} & \textbf{M}_D^T \\ \textbf{M}_D & \textbf{M}_N
        \end{pmatrix}
        \binom{\textbf{U}_L}{\textbf{U}_R^*}\textbf{U}_R^T
        \\ &=
        \textbf{U}_R
        \Big(\textbf{U}_L^T\textbf{M}_D^T\textbf{U}_R^*
        + \textbf{U}_R^\dagger \textbf{M}_D\textbf{U}_L
        + \textbf{U}_R^\dagger \textbf{M}_N\textbf{U}_R^*\Big)
        \textbf{U}_R^T
    \end{align*}
    Expanding the factors into the parenthesis, the first two terms give
    vanishing contribution by Eq.~\eqref{eq:Urel3}, while utilizing
    Eq.~\eqref{eq:Urel1}, the last one is simply $\textbf{M}_N$, so
    \beq
    \textbf{U}_R\textbf{M}\textbf{U}_R^T = \textbf{M}_N
    \,.
    \eeq
    Now Eq.~\eqref{eq:Urel3} allows us to derive
    \be
    \textbf{U}_R^\dagger \textbf{M}_N =
    \textbf{U}_R^\dagger \textbf{U}_R\textbf{M}\textbf{U}_R^T =
    (\bom{1}_6 - \textbf{U}_L^T\textbf{U}_L^*)\textbf{M}\textbf{U}_R^T =
    \textbf{M}\textbf{U}_R^T
    -\textbf{U}_L^T\textbf{U}_L^*\textbf{M}\textbf{U}_R^T
    \,.
    \label{eq:URMURT}
    \ee 
    where the second term on the right does not vanish this time.
    \section{Evaluation of the vector boson exchange diagram}\label{app:proof}
    The vector boson exchange diagrams, shown in the bottom row of \fig{fig:loop} contribute
    gauge-dependent terms to the neutrino self energy. In order to show that the 
    gauge-dependent terms cancel once contributions from all particles are considered,
    it is useful to eliminate the loop momentum from the one loop integral 
    corresponding to the vector boson exchange diagram. A decomposition to achieve 
    this was used in Ref.~\cite{Loschner:2018idt}, and we shall derive it here as well. 
    In Ref.~\cite{Weinberg:1973ua}, Eq~(4.4) contains the self energy:
    \be 
    \ri\Sigma_V(p) = -\int \frac{d^d\ell}{(2\pi)^d}\bom{\Gamma}^{\mu \dagger}
    \textbf{P}(p-\ell)\bom{\Gamma}^\nu P_{\mu\nu}(\ell,M_V^2;\xi_V),
    \ee 
    where the $6\times 6$ matrices are defined as follows. The matrix {\bf P} 
    is the fermion propagator, diagonal in the mass eigenstates,
    \be 
    \textbf{P}(p-\ell) = [(\cancel{p}-\cl)\bom{1}-\textbf{M}]^{-1}
    \,,\text{~~while~~}
    \bom{\Gamma}^\mu = -\ri e \gamma^\mu \textbf{A} 
    \,,\text{~~with~~}
    \textbf{A} = \bom{\Gamma}^L P_L + \bom{\Gamma}^R P_R.
    \ee
    In the following we shall write {\bf P} for $\textbf{P}(p-\ell)$.
    The matrix {\bf A} is self-adjoint, $\textbf{A}^\dagger=\textbf{A}$, and so is
    $\bom{\Gamma}^{L/R}$. We also introduce the abbreviation
    \be 
    \tilde{\textbf{A}} = \bom{\Gamma}^RP_L+\bom{\Gamma}^LP_R
    \,,
    \ee
    which will simplify our calculations. 
    In order to compute the loop integral easily containing the neutral vector boson 
    propagator in the neutrino self-energy loop, in this Appendix we perform tensor 
    reduction of the matrix product
    \be\label{app-eq:toexpand}
    \cl \textbf{A}^\dagger \textbf{P}\cl \textbf{A}
    \ee 
    such that the numerator factor be at most linear in the loop momentum $\ell$. 
    
    When the fermion momentum $p$ appears as $\cancel{p}$ at the extreme 
    left or right of the expression, it satisfies the Dirac equation 
    $\cancel{p} \bom{1}_6 = \textbf{M}$ (both Dirac and Majorana fermions do so), 
    thus we can replace formally $\cancel{p}$ with $\textbf{M}$. Let us first write the identity
    \bal
    \cl \textbf{A}^\dagger &= (\cl - \cancel{p} + \textbf{M})\textbf{A}^\dagger = \textbf{MA}^\dagger - (\cancel{p}-\cl)\textbf{A}^\dagger.
    \eal 
    The chiral coupling matrix $\textbf{A}$ anticommutes with the Dirac matrices
    $\gamma_\mu$, hence
    \bal
    \cl \textbf{A}^\dagger = 
    \textbf{M}\textbf{A}^\dagger -\tilde{\textbf{A}}^{\dagger}
    \Big((\cancel{p}-\cl)\bom{1} - \textbf{M} + \textbf{M}\Big) 
    &=\textbf{M}\textbf{A}^\dagger 
    -\tilde{\textbf{A}}^{\dagger}\textbf{P}^{-1} 
    - \tilde{\textbf{A}}^{\dagger}\textbf{M}\,,
    \label{app-eq:toexpand1}
    \eal
    and similarly,
    \bal 
    \textbf{P}\cancel{\ell}\textbf{A} = 
    -\textbf{A} - \textbf{P}\textbf{MA}+\textbf{P}\tilde{\textbf{A}}\textbf{M}\,.
    \label{app-eq:toexpand2}
    \eal 
    Multiplying Eqs.~\eqref{app-eq:toexpand1} and \eqref{app-eq:toexpand2}, 
    we obtain the expression \eqref{app-eq:toexpand}, and its expansion
    yields
    \be\bsp
    \cancel{\ell} \textbf{A}^\dagger \textbf{P}\cancel{\ell} \textbf{A}
    = &-\textbf{MA}^\dagger \textbf{A} 
    - \textbf{MA}^\dagger \textbf{P}\textbf{MA} 
    + \textbf{MA}^\dagger \textbf{P}\textbf{AM} 
    + \tilde{\textbf{A}}^\dagger \textbf{P}^{-1}\textbf{A} \\
    &+ \tilde{\textbf{A}}^\dagger \textbf{MA}
    - \tilde{\textbf{A}}^\dagger \tilde{\textbf{A}}\textbf{M} 
    + \tilde{\textbf{A}}^\dagger \textbf{MA} 
    + \tilde{\textbf{A}}^\dagger \textbf{M}\textbf{P}\textbf{MA} 
    - \tilde{\textbf{A}}^\dagger \textbf{M}\textbf{P}\tilde{\textbf{A}}\textbf{M}\,.
    \label{app-eq:expanded}
    \esp\ee 
    Using that $\cancel{p}\textbf{A} = \tilde{\textbf{A}}\cancel{p}$, 
    the fourth term can rearranged as
    \be\bsp
    \tilde{\textbf{A}}^\dagger \textbf{P}^{-1}\textbf{A} &= 
    \tilde{\textbf{A}}^\dagger \Big((\cancel{p}-\cancel{\ell})\textbf{1} 
    - \textbf{M}\Big)\textbf{A}
    = \half \tilde{\textbf{A}}^\dagger \cancel{p} \textbf{A} 
    + \half \tilde{\textbf{A}}^\dagger \cancel{p} \textbf{A} 
    - \tilde{\textbf{A}}^\dagger \cancel{\ell} \textbf{A} 
    - \tilde{\textbf{A}}^\dagger \textbf{MA}\\
    &= \half \cancel{p}\textbf{A}^\dagger \textbf{A} 
    + \half \tilde{\textbf{A}}^\dagger \tilde{\textbf{A}}\cancel{p} 
    -\tilde{\textbf{A}}^\dagger \cancel{\ell} \textbf{A} 
    - \tilde{\textbf{A}}^\dagger \textbf{MA}
    \esp\ee 
    The $\cancel{p}$ is on extreme left and right, hence can be replaced with \textbf{M}, giving
    \be
    \tilde{\textbf{A}}^\dagger \textbf{P}^{-1}\textbf{A} = 
    \half \textbf{M}\textbf{A}^\dagger \textbf{A} 
    + \half \tilde{\textbf{A}}^\dagger \tilde{\textbf{A}}\textbf{M}
    -\tilde{\textbf{A}}^\dagger \cancel{\ell} \textbf{A} 
    - \tilde{\textbf{A}}^\dagger \textbf{MA}\,.
    \label{app-eq:AtdP-1A}
    \ee
    Substituting Eq.~\eqref{app-eq:AtdP-1A} into Eq.~\eqref{app-eq:expanded}, we obtain
    \beq
    \cancel{\ell} \textbf{A}^\dagger \textbf{P}\cancel{\ell} \textbf{A} = 
    \textbf{M}_0 +\textbf{M}_1+\textbf{M}_2
    \label{app-eq:decomop} 
    \eeq
    where we introduced the abbreviations
    \beq
    \bsp
    \textbf{M}_0 &=-\tilde{\textbf{A}}^\dagger \cancel{\ell} \textbf{A}
    \,,\\
    \textbf{M}_1 &=
    - \half \textbf{M}\textbf{A}^\dagger \textbf{A} 
    - \half \tilde{\textbf{A}}^\dagger \tilde{\textbf{A}}\textbf{M} 
    + \tilde{\textbf{A}}^\dagger \textbf{MA}
    \label{app-eq:decomp1} 
    \esp
    \eeq
    and
    \beq
    \bsp
    \textbf{M}_2 &= - \textbf{MA}^\dagger \textbf{P}\textbf{MA} 
    + \textbf{MA}^\dagger \textbf{P}\tilde{\textbf{A}}\textbf{M} 
    + \tilde{\textbf{A}}^\dagger \textbf{M}\textbf{P}\textbf{MA} 
    - \tilde{\textbf{A}}^\dagger \textbf{M}\textbf{P}\tilde{\textbf{A}}\textbf{M}
    \\&= 
    (\textbf{MA}^\dagger - \tilde{\textbf{A}}^\dagger \textbf{M})\textbf{P}(\tilde{\textbf{A}}\textbf{M}-\textbf{MA})\,,
    \label{app-eq:decomp2} 
    \esp
    \eeq
    which correspond to constant, linear and quadratic terms in the neutrino mass 
    matrix \textbf{M}. We now discuss the contribution from each term in
    Eq.~\eqref{app-eq:decomop} separately.
    
    The first, constant term gives vanishing contribution to the loop integral 
    as it is odd in the loop momentum. The other two terms can be decomposed
    into left and right chiral pieces:
    \beq
    \textbf{M}_i = \textbf{M}_i^L P_L + \textbf{M}_i^R P_R  
    \,.
    \eeq
    Our goal is to compute the one-loop correction \eqref{eq:loopmasscorr} to the
    tree-level mass matrix of the light neutrinos. In order to obtain it, one 
    sandwiches the left handed pieces $\textbf{M}_i^L$ between the matrices $\textbf{U}_L^*$ and $\textbf{U}_L^\dagger$. Using the properties of 
    the neutrino mixing matrices of \app{app:UMrelations}, we immediately see that
    \beq 
    \textbf{U}_L^*\textbf{M}_1^L\textbf{U}_L^\dagger = 0
    \,,
    \eeq
    while lengthy computations yield
    \beq
    \bsp
    \textbf{U}_L^*\textbf{M}_2^L\textbf{U}_L^\dagger =&
    -(C^L_{V\nu\nu} - C^R_{V\nu\nu})^2\textbf{U}_L^*\textbf{M}
    \textbf{P}\textbf{MU}_L^\dagger
    \\&
    +\text{terms that do not contribute to {\bf B}}_L(0)\,.
    \label{app-eq:UL*BLULdag}
    \esp
    \eeq
    Here we outline the steps needed to reach Eq.~\eqref{app-eq:UL*BLULdag}.
    
    Firstly, in order to find the left-chiral part $\textbf{M}_2^L$, we substitute
    \textbf{A} and $\tilde{\textbf{A}}$ into Eq.~\eqref{app-eq:decomp2}. We write 
    the denominator of the fermion propagator as
    \beq
    (\textbf{P})_{ij} = 
    \delta_{ij}\frac{\slashed{p}-\slashed{\ell}+m_i}{(p-\ell)^2-m_i^2}
    \eeq
    and use the following relations for the Dirac projectors:
    \beq
    P_{L/R} (\slashed{q}+m) P_{L/R} = m P_{L/R}
    \,,\quad
    P_{L/R}(\slashed{q}+m) P_{R/L} = \slashed{q} P_{R/L}
    \,,
    \eeq
    valid for any momentum $q$ and mass $m$.
    Hence
    \beq
    \bsp
    P_{L/R} (\textbf{P})_{ij} P_{L/R} &= 
    \left((\textbf{P})_{ij} 
    - \delta_{ij}\frac{\slashed{p}-\slashed{\ell}}{(p-\ell)^2-m_i^2}\right) P_{L/R}
    \,,\\
    P_{L/R}(\textbf{P})_{ij} P_{R/L} &= 
    \delta_{ij}\frac{\slashed{p}-\slashed{\ell}}{(p-\ell)^2-m_i^2} P_{R/L}
    \,,
    \esp
    \eeq
    and therefore, we obtain
    \beq
    \textbf{M}_2^L = 
    -\textbf{M}\bg^{L\dagger}\textbf{P}\textbf{M}\bg^L 
    + \textbf{M}\bg^{L\dagger}\textbf{P}\bg^R\textbf{M} 
    + \bg^{R\dagger}\textbf{M}\textbf{P}\textbf{M}\bg^L 
    - \bg^{R\dagger}\textbf{M}\textbf{P}\bg^R\textbf{M}
    + \textbf{D}
    \label{app-eq:BL}
    \eeq
    where the last term is proportional to $(\slashed{p}-\slashed{\ell})$:
    \[
    \textbf{D} =
    [\textbf{M}(\bg^{L\dagger}-\bg^{R\dagger}) - (\bg^{L\dagger}-\bg^{R\dagger}) \textbf{M})]
    (\slashed{p}-\slashed{\ell}) [(p-l)^2\bom{1}_6-\textbf{M}^2]^{-1}
    (\bg^R\textbf{M}-\textbf{M}\bg^L)
    \,.
    \]
    Then using the matrix relations derived in Appendix A, we can compute 
    the following identities:
    \bal
    \textbf{U}_L^*\textbf{M}\bg^{L\dagger} &= 
    -C_{V\nu\nu}^R\textbf{U}_L^*\textbf{M}
    \,,\quad
    \textbf{U}_L^*\bg^{R\, \dagger} =
    -C^L_{V\nu\nu}\textbf{U}_L^*
    \,,\\
    \bg^R\textbf{MU}_L^\dagger &= 
    C^R_{V\nu\nu}\textbf{MU}_L^\dagger 
    \,,\qquad\;
    \bg^L\textbf{U}_L^\dagger = 
    C^L_{V\nu\nu}\textbf{U}_L^\dagger
    \,.
    \eal 
    Finally sandwiching Eq.~\eqref{app-eq:BL} gives us
    \beq
    \bsp
    \textbf{U}_L^*\textbf{M}_2^L  \textbf{U}_L^\dagger &= 
    -(C^L_{V\nu\nu} - C^R_{V\nu\nu})^2\textbf{U}_L^*\textbf{M}\textbf{P}\textbf{MU}_L^\dagger +
    \textbf{U}_L^*\textbf{D}\textbf{U}_L^\dagger
    \,.
    \esp
    \eeq
    As mentioned, the last term is proportional to $(\slashed{p}-\slashed{\ell})$, 
    but only the term with $\slashed{\ell}$ contributes to $\textbf{B}_L(p=0)$. 
    That piece, being an odd function of $\ell$, vanishes upon integration, 
    which completes the proof of Eq.~\eqref{app-eq:UL*BLULdag}.
    
    The charged vector bosons $W^\pm$ also contribute to the neutrino self-energy. 
    The corresponding Feynman rules are
    \be 
    \bom{\Gamma}^\mu_{W^-\bar{\ell}\nu} 
    = -\ri e\gamma^\mu\bom{\Gamma}^L_{W^-\bar{\ell}\nu}  P_L,
    \quad 
    \bom{\Gamma}^\mu_{W^+\nu \ell} 
    = -\ri e\gamma^\mu\bom{\Gamma}^L_{W^+\nu \ell}P_L.
    \ee 
    where 
    \be 
    \bom{\Gamma}^L_{W^-\bar{\ell}\nu}  = C_{W \ell\nu} 
    (\textbf{U}_L^\ell \textbf{U}_L)_{ij}
    \ee 
    with $\textbf{U}_L^\ell$ being the charged lepton mixing matrix and 
    $\bom{\Gamma}^L_{W^-\bar{\ell}\nu}= (\bom{\Gamma}^L_{W^+\nu \ell})^\dagger$. 
    The charged vector boson contribution to \eqref{eq:loopmasscorr} is proportional 
    to $\textbf{U}_L^* \textbf{M} \textbf{U}_L^\dagger$, which vanishes identically 
    as shown in \app{app:UMrelations}.
    \bibliographystyle{unsrt}
    \bibliography{ArxivVersion_Final.bib}
\end{document}